# The Design and Implementation of a Virtual Statistical Computing Lab to Teach R Coding to Introductory Statistics Students

**Sayed A Mostafa**[1,*], **Tamer M Elbayoumi**[1], **Seongtae Kim**[1], **and Oluwatobi Akinbode**[1]

[1]Department of Mathematics & Statistics, North Carolina A&T State University, Greensboro, NC, USA
***Corresponding Author**: sabdelmegeed@ncat.edu

**Abstract**: Motivated by national calls for computationally enriched, data-centric instruction across the statistics curriculum, this study investigates the design, implementation, and impact of a Virtual Statistical Computing Lab (VSCL) integrated into an introductory statistics course at a medium-sized minority-serving university in the USA. The redesigned course embedded R-based coding through two virtual lab formats: Design I (a static Posit Cloud environment) and Design II (an interactive *learnr*-based interface). Using a quasi-experimental design across three instructional formats—traditional (no lab), Design I, and Design II—we evaluated students' conceptual learning gains, levels of data science (DS) readiness, and DS aspirations. The results indicated significant learning gains across all groups, with the highest gains observed in Design II. Students in both VSCL formats achieved greater gains in DS readiness than the traditional group, with Design II again yielding the largest gains across the demographic subgroups. Conversely, DS aspirations remained low or declined, suggesting a gap between skill acquisition and long-term interest. These findings highlight the value of structured, interactive computing environments in supporting statistical reasoning and building confidence in modern data tools. They also point to the need for intentional curricular bridges and career mentoring to help students translate early computing exposure into sustained academic and professional pathways in statistics and data science.

## 1. INTRODUCTION

Statistics educators have been especially active in recognizing that tomorrow's graduates must be comfortable not only in interpreting quantitative results but also in generating them using modern computational tools. As personal computing devices and cloud resources become increasingly powerful and affordable, integrating technology into introductory statistics courses has become indispensable, enabling instructors to demonstrate abstract concepts such as sampling variability, simulation, and probabilistic reasoning in ways that traditional lecture slides alone cannot.

One effective strategy is the inclusion of structured, hands-on laboratory sessions that run parallel to traditional lectures, providing students with repeated opportunities to wrangle real data, visualize patterns, and test hypotheses in a guided setting. These labs often rely on shared digital platforms—termed *virtual computing labs* by Li et al. (2009)—that provide students with identical software configurations and datasets. These environments eliminate the logistical challenges of individual installations by allowing instructors to focus on conceptual understanding and student engagement rather than technical troubleshooting.

Delivering rich curricular experiences, whether in traditional classrooms or fully online, requires intentional instructional design. However, many instructors have limited prior exposure to the pedagogies and technical logistics that make virtual labs effective. Identifying best practices and digital toolkits is therefore essential if we aim to preserve—and elevate—the depth of student learning. A central part of this effort involves promoting a view of data modeling not as a fixed set of formulas, but as an iterative, inquiry-driven process for extracting insight from complexity. To support this shift, we advocate embedding a Virtual Statistical Computing Lab (VSCL) into statistics and data science (DS) coursework. A VSCL offers students repeated practice in coding environments where they simulate multivariate data, construct models, and develop the computational routines that support business solutions and scientific discovery. These skills are inherently computational and cannot be fully developed through lectures alone.

Nolan and Temple (2010) urged the discipline to treat programming fluency on par with mathematical rigor, calling for a curricular overhaul. Subsequent recommendations by the American Statistical Association's 2016 undergraduate guidelines and the National Academies' 2018 consensus report on data science education reinforce the same message: undergraduates must gain mastery in reproducible workflows, principled data wrangling, and modern statistical software. Integrating VSCL experiences supports this mandate by providing structured, hands-on opportunities to engage with the full statistical reasoning cycle—from data acquisition to transparent, shareable analysis.

Although many instructors remain concerned about the overcrowded nature of the undergraduate statistics syllabus, growing evidence indicates that embedding computational components is not an optional "add-on," but a critical evolution. Reforms emphasizing simulation-based inference and early exposure to DS tools are steadily reshaping entry-level courses, encouraging instructors to replace traditional instruction with experiential, code-driven learning (Tintle et al., 2015; Kaplan, 2018). For instance, Baumer et al. (2020) paired interactive DataCamp lessons with traditional lectures to scaffold learning through modular content, custom R packages, and dashboards for real-time progress-tracking. Similarly, Gehrke et al. (2021) advocated for a workflow-centered approach in which students acquire, clean, visualize, and model messy datasets, ultimately translating the results into plain language. Their framework emphasizes four core habits: recognizing when evidence is needed, extracting signals from noise, acknowledging uncertainty, and constructing models that connect data to real-world problems. Framing statistical computing in this way shifts it from a perceived curricular burden to an engine that powers statistical literacy and critical inquiry.

## 1.1. Coding in Introductory Statistics

As analysts work with increasingly complex datasets and evolving tools, the demand for transparent and reproducible reporting has become central to statistical practice. Meeting this challenge in introductory courses involves three essential components: selecting an accessible programming language, ensuring that the code is both correct and readable, and integrating analysis with a coherent narrative. Diez et al. (2019) show how open-access resources can integrate theory, simulation, and real data through cohesive R-based exercises. Baumer et al. (2014) emphasized how R Markdown fosters literate programming by uniting code, output, and explanation, thus improving both engagement and clarity. Bray et al. (2018) extend this approach with the *infer* package, which aligns with the *tidyverse* and streamlines inference syntax. More recently, Fergusson (2024) introduced a sequence of *learnr*-based lab modules that guide students from scaffolded exploration to independent analysis. These innovations frame statistical computing as a unified process that involves coding, reasoning, and communication.

One of our earliest redesign decisions addressed the "double wall" faced by beginners: fear of syntax and uncertainty about coding's relevance. Doehler and Taylor's (2015) field study demonstrated that this anxiety can be reduced through short, scaffolded code snippets integrated into traditional instruction, which gradually expand to full scripts. Following this model, each lab began with a runnable template, introduced the syntax incrementally, and concluded with an applied analysis using a real dataset. This helped students gain confidence by framing coding as a purposeful and supportive learning tool.

We selected R not only for its open-source nature, but also because of its widespread use in statistical and data science communities. Çetinkaya-Rundel and Ellison (2021) emphasized that incorporating hands-on R coding into introductory coursework helps students develop analytical workflows, fosters reproducibility, and provides technical skills for students. To support these goals, our implementation included screencasts on data import, wrangling, and visualization, embedded directly into the lecture materials. Students accessed live RStudio Cloud environments, preconfigured with the necessary packages, via hyperlinks in the course notes. Weekly assignments asked students to replicate textbook figures using *ggplot2* and explain each code layer in plain language. Over time, they became more confident in reading and adapting unfamiliar code, which is an essential skill for working with open-source tools and research supplements. This emphasis on transferability, self-paced learning, and no-cost software aligns with equity goals and prepares students for research settings that increasingly rely on scripted and reproducible workflows.

Kaplan (2007) argues that computational thinking is essential for understanding randomness. In our VSCL, students worked through live code chunks to simulate and test assumptions, developing what Kaplan called "algorithmic intuition." Rivera et al. (2019) extended this idea by pairing civic datasets with reproducible R scripts. Inspired by this model, we included editable R Markdown files with embedded checkpoints to help students trace the connection between the code and output. These habits—transparent scripting, version control, and reproducibility—mirror the best practices in professional data work.

To bridge classical inference and modern data science methods, students progressed from base R explorations to *tidyverse* pipelines, culminating in machine learning workflows such as multivariate regression. Resources such as the *mosaic* package (Lübke et al., 2019) and Dalgaard's (2008) conceptual guides supported this progression. The VSCL ensured that all students, regardless of the device or operating system, worked in a uniform, preconfigured environment, which is a key condition for equitable access and minimal troubleshooting. By semester's end, students were not only able to conduct statistical inference and interpret p-values, but also confident in using wrangling functions such as *filter(), select()*, and *mutate()* to build reproducible, meaningful analyses.

## 1.2. Computing Labs for Introductory Statistics

Recent curriculum frameworks increasingly affirm that authentic computing labs—whether physical or browser-based—are vital for helping students transition from rote procedures to genuine data investigation. The GAISE College Report (Carver et al., 2016) advocates for technology-rich activities that immerse students in empirical work, emphasizing analysis and interpretation over mechanical calculations. Our redesign responds directly to this charge: every lecture is paired with a lab session where core concepts are immediately applied using executable R scripts or applet-based simulations. This pairing reinforces the idea that statistical concepts take shape in code, not just symbols.

This approach aligns with Cabilio and Farrell's (2001) studio-lab model, which supplements lectures with real-time data manipulation and simulation to foster active learning. Guardiola et al. (2010) compared three instructional models—lecture only, lecture with an uncoordinated lab, and lecture with a coordinated lab—and found that students in the coordinated lab condition performed significantly better.

Virtual labs extend these benefits by decoupling instruction from time and location constraints. Lane's Rice Virtual Lab (1999) was a pioneer in this space, allowing users to interactively explore sampling distributions through browser-based simulations. Today's virtual labs build on that foundation by embedding full R or Python environments in the cloud, enabling students to move seamlessly from GUI-based tools to script-based analysis within a single platform. This model scaffolds coding skill development while preserving statistical rigor. Standardized cloud containers also allow instructors to diagnose issues, push updates, and offer real-time support with minimal configuration variability.

Hybrid models—combining synchronous video sessions with cloud computing—merge the immediacy of live dialogue with the flexibility of online tools. Dixson (2010) and King (2014) emphasized that consistent instructor presence via video, mini-lectures, and screencasts enhances engagement and improves outcomes. Albert et al. (2020) found similar effects in MOOC settings, where integrating animations, R code, and visualizations in lieu of live lecture improved retention. In our VSCL, lab recordings are preserved for asynchronous access, offering flexibility and review opportunities for all students. Each lab begins with a substantive reading and ends with a practical assignment rooted in code output and visual evidence, in line with GAISE's warning that technology should serve, not overshadow, statistical thinking.

## 1.3. Static versus Interactive Coding

Provisioning students with a static desktop IDE (e.g., RStudio) or a locked-down Posit Cloud project offers a stable, uniform environment. Every student accesses the same directory and package versions, which ensures reproducible performance across machines. This consistency is critical for more advanced tasks such as feature engineering, multi-table joins, and background modeling. Sihler et al. (2024) showed that such setups support high-performance tasks like memory mapping and GPU acceleration. Static environments also allow for integration with advanced linting tools that surface coding flaws, helping students develop best practices through guided revision (Liu et al., 2024; Haindl & Weinberger, 2024).

By contrast, interactive environments—such as those built with the *learnr* package—embed code, quizzes, and hints into the learning narrative, creating an engaging sandbox. Marcinkowska and Roszak (2024) paired *learnr* with dynamic Shiny dashboards to help students explore sampling variability interactively. Tools like *infer*, *gradethis*, and *learnrhash* (Rundel, 2021) support auto-scoring and progress tracking, allowing personalized feedback at scale (Stoudt et al., 2022).

Neither paradigm alone suffices. A hybrid approach, beginning with interactive tutorials and transitioning to static scripting, supports both exploration and production-level skill development. Studies by Woodard & Lee (2021) and Tintle et al. (2018) show that computation embedded early in coursework narrows achievement gaps and promotes conceptual gains. As students mature, interactive feedback gives way to literate scripting and workflow automation, equipping them for professional and academic settings (Altman & Krzywinski, 2015; Xie et al., 2020; Grolemund & Wickham, 2017).

## 1.4. Research Questions

To measure the effectiveness of two virtual computing lab formats in the redesigned introductory statistics course, this study addresses the following research questions:

**RQ1:** To what extent does the integration of an R-based virtual computing lab in introductory statistics improve students' conceptual statistical learning gains?
**RQ2:** To what extent does the integration of an R-based virtual computing lab in introductory statistics enhance students' self-perceived readiness for data science tasks?
**RQ3:** What impact does the integration of an R-based virtual computing lab in introductory statistics have on students' aspirations to pursue further studies or credentials in data science?
**RQ4:** How do static and interactive computing lab designs compare in their effects on these student outcomes relative to a traditional course design?

The remainder of this paper is organized as follows: Section 2 describes the study setting and outlines the three comparative instructional designs—traditional, Posit cloud–based VSCL (Design I), and *learnr*-based VSCL (Design II). Section 3 details the assignment mechanism, data collection, measurement instruments, and the statistical analysis framework. Section 4 presents the results regarding conceptual learning gains, data science readiness, and aspirations, including psychometric validation of measures and sensitivity analyses for missing data. Section 5 concludes with a discussion of key findings, implications for instructional design, and future research directions.

# 2. BACKGROUND

## 2.1. Study Setting

The “Introduction to Probability & Statistics” course in this study is an algebra-based introductory statistics course offered at a medium-sized minority-serving university in the United States. The course serves a diverse group of students, with approximately 46% from STEM fields and 54% from non-STEM disciplines. Notably, the majority of students come from underrepresented groups in the fields of Statistics and Data Science–about 82% are African Americans, and 69% are female. Typically, seven sections of the course are offered during each of the Fall and Spring semesters, with 30 to 40 students in each section. In the summer of 2021, we received a grant from the National Science Foundation (NSF) to redesign the course to boost students’ statistical competencies and to motivate and prepare them for further data science education and career opportunities. The redesigned course, named “DS-infused Intro Stats,” was developed in Fall 2021 and piloted in Spring 2022. The course structure was revised to integrate data science concepts into traditional statistical topics, offering students practical, hands-on experience in data analysis, coding, and interpretation. This infusion of data science tools was intended to provide a modernized learning experience, making the curriculum more relevant to the growing demands in the data-driven job market. The project is detailed further at https://introtostatncat.github.io, and the course redesign is part of a broader effort to address disparities in the field by preparing underrepresented students for future success in statistics and data science.

## 2.2. Traditional Intro Stats Course (No Coding)

The traditional Intro Stats course at the institution under study involves three hours of weekly lectures. Delivered in what is commonly referred to as a “traditional manner,” the course includes instructor-led sessions where definitions, formulas, and procedural steps are presented through the use of whiteboards, PowerPoint slides, or a combination of the two. Practice problems are solved in class to guide students’ understanding, while calculations are primarily performed using calculators and statistical distribution tables. This instructional approach aligns with what the literature describes as a “consensus” introductory statistics course (Cobb, 2015).

The course employed “OpenIntro Statistics 4th Edition” as its textbook and covered a standard range of topics (Diez et al., 2019). These topics included an introduction to data basics, data collection methods, summarizing numerical and categorical data, and an introduction to probability. Furthermore, the curriculum encompassed the distributions of random variables, sampling distributions, one- and two-sample inference for means and proportions, and an introduction to linear regression. To provide additional support, an optional one-hour recitation session was made available for students enrolled in the traditional course sections. The recitation session helped maintain a balance in weekly time commitments between students in traditional sections and those participating in the redesigned, computationally-infused sections, which are described next.

## 2.3. Intro Stats with a Virtual Computing Lab

### *Design I (Posit-Cloud-Based Labs)*

Under Design I, the introductory statistics course maintained the same weekly three-hour class structure and textbook as the traditional course, while incorporating additional data science concepts and computational tools (see Figure 1). A significant enhancement to the course was the introduction of a one-hour virtual statistical computing lab, held weekly via Zoom and utilizing the Posit Cloud platform. During these lab sessions, students engaged in R coding exercises that reinforce the statistical concepts and methods discussed in class. These sessions were facilitated by a graduate teaching instructor with assistance from graduate assistants, ensuring students received guidance and support as they worked through the tasks.

The weekly lab sessions utilized static HTML-based lab descriptions adapted from the OpenIntro Statistics labs available at https://www.openintro.org/book/os. These labs introduced students to the *tidyverse* ecosystem, specifically focusing on data transformation and summarization. For instance, students used the *summarise*() function to compute measures of central tendency (mean, median) and spread (standard deviation, interquartile range) using real-world datasets like the *nycflights13* dataset. Further exercises guided learners to create filtered datasets and compute grouped statistics using *group_by*(), fostering an early understanding of how to describe and compare distributions across different categories (e.g., carrier delays). Visual examples of the static lab interface and exercise structure are provided in Appendix A (Figure A1).

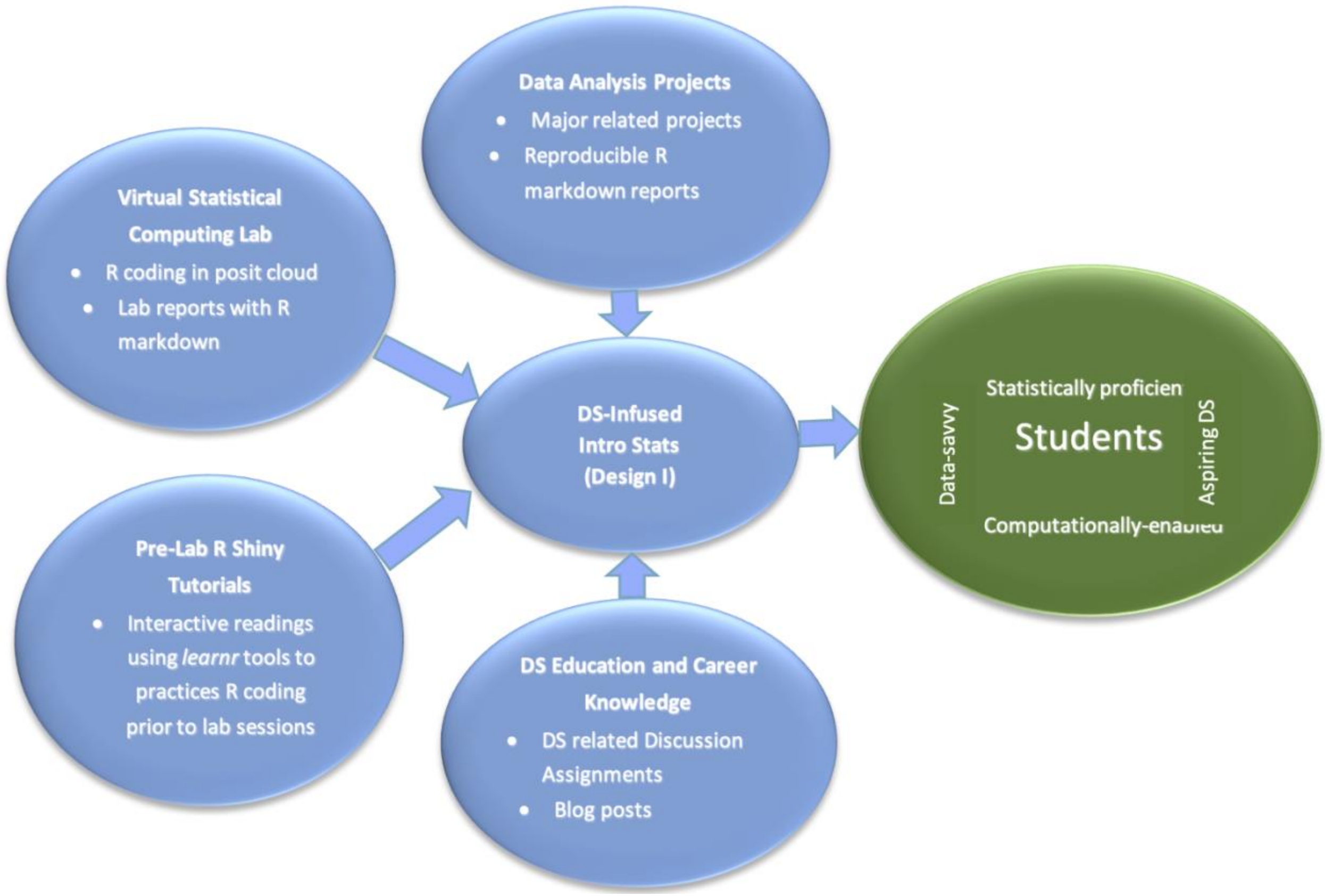

Figure 1: Course design I for Intro Stats with Posit Cloud-based computing labs using R markdown.

To prepare for these labs, students completed for-grade pre-lab tutorials. These were developed using the *learnr* package and hosted on a Shiny server, with many modules adapted from the interactive notebooks of Gilbert (2022). These tutorials utilized the *gradethis* functionality (Aden-Buie et al., 2025) to provide automated feedback on foundational tasks, such as identifying outliers and comparing measures of central tendency. This approach ensured students had hands-on experience with the R interface before the synchronous Zoom session. After each lab, students submitted a reproducible PDF report generated from an R Markdown template preloaded in their Posit Cloud workspace (see Appendix A, Figures A2 and A3, for the tutorial interface and report templates).

In addition to the computing labs, each student conducted a semester-long data analysis project using a dataset aligned with their specific academic major. The project was scaffolded into three parts, each building upon the skills and concepts introduced throughout the semester.

- Data Wrangling and Exploration: Students used *tidyverse* to clean their datasets and generate initial summary statistics to formulate research questions.
- Statistical Inference: Using the *infer* package (Bray et al., 2018), students conducted hypothesis tests and constructed confidence intervals (e.g., using *t_test*()), while visualizing results through boxplots, histograms, and violin plots.
- Multivariate Modeling: The final phase covered pairwise correlations, scatterplots, and regression modeling, culminating in a comprehensive R Markdown project report.

This curriculum aligns with recommendations by Hardin et al. (2015) and Horton & Hardin (2021) for integrating computing early and frequently in statistics education. The lab structure promotes active learning and real-data analysis using technology, as recommended by the Guidelines for Assessment and Instruction in Statistics Education (GAISE) College Report (GAISE, 2016).

To further engage students and spark interest in statistics and data science careers, the course also incorporated asynchronous online discussion forums highlighting the role of data in solving real-world problems, along with curated blog posts on educational pathways and trends in the data science job market.

### *Design II (learnr-based Interactive Labs)*

To further enhance the DS-Infused Intro Stats course and alleviate computational anxiety, Design II introduced a hands-on interactive approach utilizing the *learnr* package (see Figure 2). This transition aimed to boost student confidence by moving away from static descriptions toward a "live" coding environment within the browser. Virtual labs were transformed into interactive R modules that provided immediate, automated feedback, allowing students to identify and correct syntax or logic mistakes in real-time (see Appendix A, Figure A4 for an example interactive lab with exercises). As in Design I, dedicated lab time was allocated to ensure that students could make consistent progress on their semester-long data analysis projects. At the end of the lab session, students can check their score and generate a hash code, which is then submitted for grading in the Blackboard LMS. To facilitate seamless and secure grading, the VSCL utilized this hash code as a unique, encrypted digital signature that encodes the student's progress and final score within the *learnr* environment. By copying and pasting this code into the LMS, students provide a tamper-proof verification of their lab completion. This mechanism allows the instructor to verify authentic work and maintain data integrity, while preserving the lightweight, browser-based nature of the interactive environment. A simple R script was created to process the hash codes using the *learnrhash* package and generate a spreadsheet with students' names and scores, which is then uploaded back to the LMS (the submission interface and auto-scoring script are illustrated in Appendix A, Figures A5 and A6).

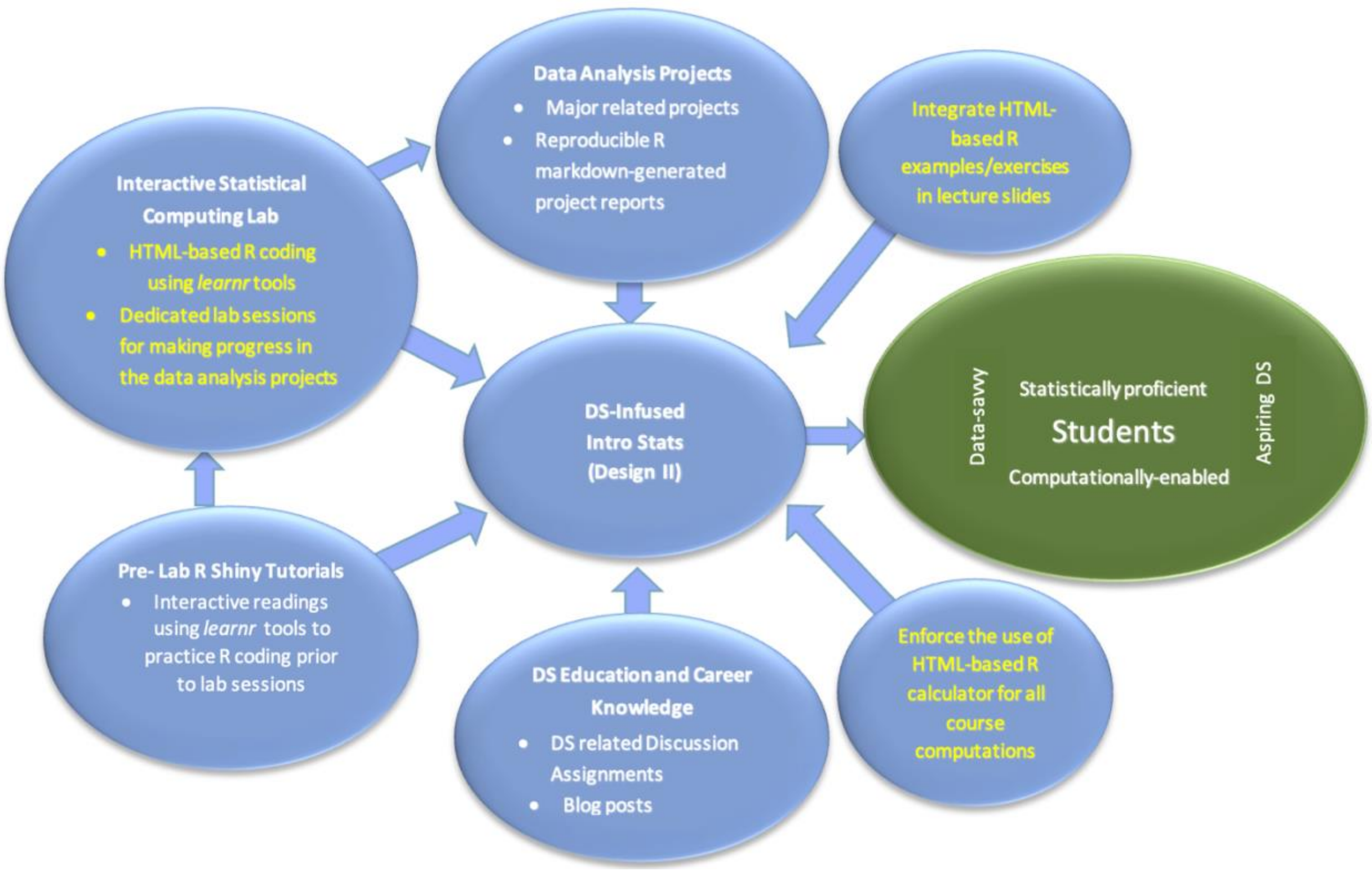

Figure 2: Course design II for Intro Stats with <u>interactive</u> computing labs using *learnr.*

To maintain pedagogical consistency, interactive R examples and exercises were also integrated directly into the lecture notes and presentation slides (see Appendix A, Figure A7). This integration allowed students to test statistical methods during the lecture, bridging the gap between theoretical instruction and practical implementation and promoting active participation. Furthermore, an HTML-based R calculator (https://introtostatncat.shinyapps.io/R_calculator/) was introduced for all course computations to standardize the computational approach across the curriculum. This consistent use of R throughout the course reinforced the computational skills developed in the labs, preparing students for further data science-infused coursework.

# 3. METHODS

This section details the assignment mechanism, measurements, data, and statistical framework used to address the research questions outlined in Section 1.4. Specifically, we describe the instruments used to assess students' statistical conceptual learning gains, data science readiness, and aspirations, alongside institutional records used to capture demographics and academic history. We also present descriptive statistics to summarize the sample characteristics. Lastly, we detail the statistical methods used to analyze changes in target student outcomes and to explore potential driving factors.

## 3.1. Assignment Mechanism

The study employed a quasi-experimental, section-level assignment mechanism across two academic semesters to evaluate the impact of the VSCL on student outcomes in introductory statistics. Students enrolled in course sections through the university's standard registration system. All course sections were listed with identical credit hours (three hours of lecture and one laboratory hour per week), course numbers, and catalog descriptions. Importantly, students were not informed which instructional format (traditional, Design I, or Design II) a given section would use at the time of registration. As a result, students could not self-select sections based on exposure to R, computing, or the VSCL. Instructional format was therefore assigned at the section level, creating a clustered quasi-experimental design rather than individual randomization. In Fall 2022, four sections were offered: two traditional sections and two Design I (static VSCL) sections. Two instructors (A and B) each taught one traditional and one Design I section. In Spring 2023, six sections were offered: two traditional sections and four Design II (interactive VSCL) sections. Instructors A and B again taught both treatment and control sections, with A teaching one Design II and one traditional section, while B teaching two Design II sections and one traditional section. A third instructor (C) taught one additional Design II section.

This assignment mechanism approximates an as-if random assignment at the section level, providing a credible basis for estimating the impact of the two VSCL formats relative to traditional instruction, for the following reasons. First, students were blind to instructional format when enrolling, so enrollment decisions could not be driven by preferences for or against computing. Second, all sections were structurally equivalent in contact hours, credits, and official descriptions of the registrar's office. Third, instructors A and B each taught both VSCL and traditional sections in both semesters, minimizing confounding between treatment and instructor characteristics. Consequently, any remaining differences between groups are most plausibly attributable to the VSCL implementation rather than systematic student sorting or instructor effects.

## 3.2. Data and Measurements

To evaluate the impact of the VSCL on student outcomes, we collected and integrated three primary data sources: 1) self-report survey data, 2) conceptual content assessments, and 3) institutional records.

The attitudes survey included questions assessing students' perceived competencies in statistical reasoning, data summarization, modeling, and reporting, as well as DS aspiration questions on plans to pursue data science coursework or credentials (see Table 1 for the question lists). The survey was administered via Qualtrics both at the beginning and end of the semester. Engagement with the survey was high and consistent across cohorts, with pre-survey completion rates exceeding 94% for the Traditional, Design I, and Design II groups. While participation moderated slightly by the end of the term, post-survey completion remained consistent across all three formats, ranging from 69.6% to 76.1% (see Table 2).

Table 1. Data science readiness and aspirations survey questions.

| | |
|---|---|
| **Readiness Questions** (6-point Likert scale: 0 = strongly disagree to 1 = strongly agree) | I can summarize data sets with summary statistics and graphics using the RStudio software. |
| | I can perform basic statistical inference using the RStudio software. |
| | I can perform basic statistical modeling (linear and/or logistics regression). |
| | I can create reproducible data analysis reports using the R Markdown software. |
| | I am adequately prepared to apply statistical and data-analytical techniques and/or tools to study a given topic. |
| **Aspiration Questions** (No, Unsure, or Yes) | Do you plan to take Data Science course(s) during your undergraduate program or during your graduate study (if you plan to do graduate studies)? |
| | Do you plan to complete a certificate in Data Science during your undergraduate program or during your graduate study (if you plan to do graduate studies)? |
| | Do you plan to complete a minor in Data Science during your undergraduate program or during your graduate study (if you plan to do graduate studies)? |
| | Do you plan to complete a degree in Data Science during your undergraduate program or during your graduate study (if you plan to do graduate studies)? |

To evaluate students' conceptual understanding of statistics, we used the Comprehensive Assessment of Outcomes in Statistics (CAOS) test. The CAOS is a validated 33-item multiple-choice instrument originally developed by delMas et al. (2007) and revised by Tintle et al. (2018). It assesses core statistical concepts, including variability, sampling, graphical reasoning, and inference. The test was administered as a pre-test and post-test during class (~40 minutes each), with participation incentives provided. Pre-test completion rates were high, particularly within the experimental cohorts (VSCL groups), with 98.6% of Design I and 97.3% of Design II students participating, compared to 77.0% in the Traditional group. This high level of initial engagement was sustained through the post-test, where completion rates were 81.7% for Design I, 91.1% for Design II, and 82.3% for the Traditional group. Overall, participation rates suggest good engagement with the CAOS pre/posttest across the three groups.

Student institutional data included demographic information (Gender, residency status, and Pell Grant eligibility as a proxy for socioeconomic status), academic background (cumulative GPA, final course grade, and Advanced Placement (AP) Statistics credit), and attendance, which was obtained from the university's office of institutional research. All data collection was approved by the university's Institutional Review Board (Protocol # HS19-0108).

The final analytic sample included participants who completed at least one of the pre-/post-content tests (CAOS) and at least one of the pre-/post-surveys. Out of 362 eligible students, 296 (81.8%) met these inclusion criteria. The sample was distributed across the three instructional formats as follows: Traditional ($N = 113$; 4 sections), Design I ($N = 71$; 2 sections), and Design II ($N = 112$; 4 sections).

Table 2 presents the demographic and academic profile of the participants, grouped by course design. Overall, the three groups demonstrated somewhat comparable baseline characteristics. Notably, the Design II group exhibited the highest post-test completion rate (91.1%), while the Traditional group showed slightly higher pre-course STEM enrollment (67.3%).

Table 2. Participant demographics, academic background, and assessment completion rates by course design.

| Variable | Value | Traditional ($N$ = 113) | Design I ($N$ = 71) | Design II ($N$ = 112) |
|---|---|---|---|---|
| **Assessment Completion** | | | | |
| Pretest Completed | Yes (%) | 87 (77.0) | 70 (98.6) | 109 (97.3) |
| Posttest Completed | Yes (%) | 93 (82.3) | 58 (81.7) | 102 (91.1) |
| Presurvey Completed | Yes (%) | 107 (94.7) | 67 (94.4) | 106 (94.6) |
| Postsurvey Completed | Yes (%) | 86 (76.1) | 52 (73.2) | 78 (69.6) |
| **Demographics** | | | | |
| Gender | Male (%) | 38 (33.6) | 26 (37.1) | 45 (40.2) |
| Pell Eligibility | Yes (%) | 79 (69.9) | 61 (87.1) | 92 (82.1) |
| Residency | Out-of-State (%) | 42 (37.2) | 22 (31.4) | 37 (33.0) |
| Rural | Yes (%) | 15 (13.3) | 16 (22.9) | 19 (17.0) |
| **Academic Background** | | | | |
| STEM Major | Yes (%) | 76 (67.3) | 49 (70.0) | 68 (60.7) |
| Pre-course GPA | ≥ 3.0 (%) | 53 (52.5) | 35 (55.6) | 70 (62.5) |
| AP Statistics | Yes (%) | 33 (30.8) | 20 (29.9) | 27 (25.5) |
| **Course-Level Metrics** | | | | |
| Attendance Rate | Mean (SD) | 82.1 (15.4) | 78.1 (20.9) | 82.9 (15.9) |
| Final Course Grade | A (%) | 26 (23.0) | 15 (21.1) | 21 (18.8) |
| | B (%) | 27 (23.9) | 12 (16.9) | 39 (34.8) |
| | C (%) | 30 (26.5) | 23 (32.4) | 38 (33.9) |

## 3.3. Statistical Analysis

To evaluate changes in student outcomes, including conceptual learning gains and data science (DS) readiness, while accounting for within-student dependence and minimizing data loss due to incomplete pre–post testing, two complementary analytic strategies were employed. First, within each instructional design, partially overlapping t-tests (Derrick & White, 2022) were conducted to assess pre- to post-course changes in CAOS and data science readiness scores. This approach incorporates both paired observations (students with complete pre–post data) and unpaired observations (students with only one measurement), allowing all available data to contribute to the estimation of within-group change without the listwise deletion required by standard paired t-tests. Second, to formally compare learning gains across instructional formats while accounting for the dependency between repeated measurements, a two-way repeated measures analysis of variance (ANOVA) was conducted, with time (Pre, Post) specified as a within-subject factor and instructional design (Traditional, Design I, Design II) specified as a between-subjects factor. The primary inferential focus of this model was the time × instructional design interaction, which tests whether pre–post changes differed across instructional formats and provides the principal evidence of differential instructional effects. Partially overlapping t-tests were conducted using the partover.test() function from the "*partiallyoverlapping*" R package (Derrick, 2018). Repeated measures ANOVA models were estimated using the anova_test() function from the "*rstatix*" package with Type III sums of squares. Effect sizes were reported using generalized eta squared (GES), and statistical significance was evaluated at a 5% significance level.

Multivariable linear regression was also conducted for CAOS post-test and DS readiness post-test scores to examine the association of the instructional design with student learning gains and DS attitudes while controlling for various covariates. Each model included pre-test scores, instructional design, student demographic and academic background, class attendance, and course grade. To address the missing values in both response and explanatory variables, we used the multiple imputation by chained equations (MICE) technique from the R package "*mice*" (Van Buuren and Groothuis-Oudshoorn, 2011). MICE imputes incomplete variables using a series of regression models that are updated iteratively. This approach is flexible and accommodates various data types: linear regression is typically used for continuous variables, whereas logistic regression is applied to binary variables. The number of multiple imputations used is $m = 50$ for modeling both the CAOS post-test score and the DS readiness post-test score. According to the rule-of-thumb, suggesting $m$ to be at least equal to the percentage of missingness (White et al., 2011), $m = 50$ is sufficient given the highest percentage of missingness in any variable within the two models being 14.5% and 27%, respectively. In this analysis, the imputation models included all variables used in the subsequent regression: pre-course CAOS score/DS readiness score, course design, gender, residency status, rural background, PELL eligibility, STEM major status, AP Statistics experience, binary GPA category, class attendance, and final course grade (A/B, C, or below C). Missing values in these predictors were imputed using the predictive mean matching (PMM) method, a semi-parametric method that preserves the original distribution of the data by matching missing values to observed values with similar predicted values (van Buuren & Groothuis-Oudshoorn, 2011). Linear regression models were then fitted separately to each imputed dataset, with the post-test CAOS score or DS readiness score as the outcome. The results were subsequently pooled using Rubin's formula to obtain combined estimates, standard errors, and confidence intervals that account for the uncertainty introduced by missing data (Rubin, 1987). Model performance was assessed by averaging $R^2$ and adjusted $R^2$. A Wald test was used to assess the joint significance of the predictors. This process allowed us to retain all cases in the analysis and reduce bias that could result from the listwise deletion of missing values.

It should be noted that the use of multiple imputation via MICE relies on the assumption that the data are Missing at Random (MAR). Under MAR, the probability that a data point is missing depends only on the observed data and not on the missing values themselves (Rubin, 1976). To evaluate the robustness of the findings to this treatment of missing data, a sensitivity analysis was conducted by re-estimating all regression models using the complete case data only. These models were specified identically to the imputed-data analyses and included the same set of predictors. Estimates from the complete case analyses were then compared with those obtained from the multiply imputed datasets in terms of coefficient magnitude, direction, and statistical significance. All results are reported in the next section.

# 4. RESULTS

This section presents the analysis results addressing the study's primary research questions regarding the efficacy of VSCL course designs for introductory statistics. The results are organized into three subsections: (1) conceptual learning gains measured by the CAOS assessment, (2) perceived DS readiness derived from the pre/post surveys, and (3) shifts in DS aspirations. Differences are explored across course designs, demographic subgroups, and academic background factors.

## 4.1. Students' Conceptual Learning Gains

The first research question (RQ1) evaluates the impact of the proposed VSCL designs on students' conceptual understanding of statistics. The following empirical results explain whether the integration of R coding and virtual computing labs enhances or hinders the mastery of core statistical concepts.

### *Trends and Subgroup Variations in Conceptual Mastery*

Figure 3a illustrates the distribution of CAOS scores across time (pre-test vs. post-test) and instructional format. Across all three instructional formats, Traditional, Design I, and Design II, students demonstrated measurable improvements in conceptual understanding. While the Traditional group began the semester with a slightly higher pre-test baseline, Design II yielded the highest post-test scores (a mean of 47.4), followed by the Traditional group (46.3) and Design I (44.8). Analysis of demographic and academic subgroups (Figures 3b–3h) revealed several notable trends:

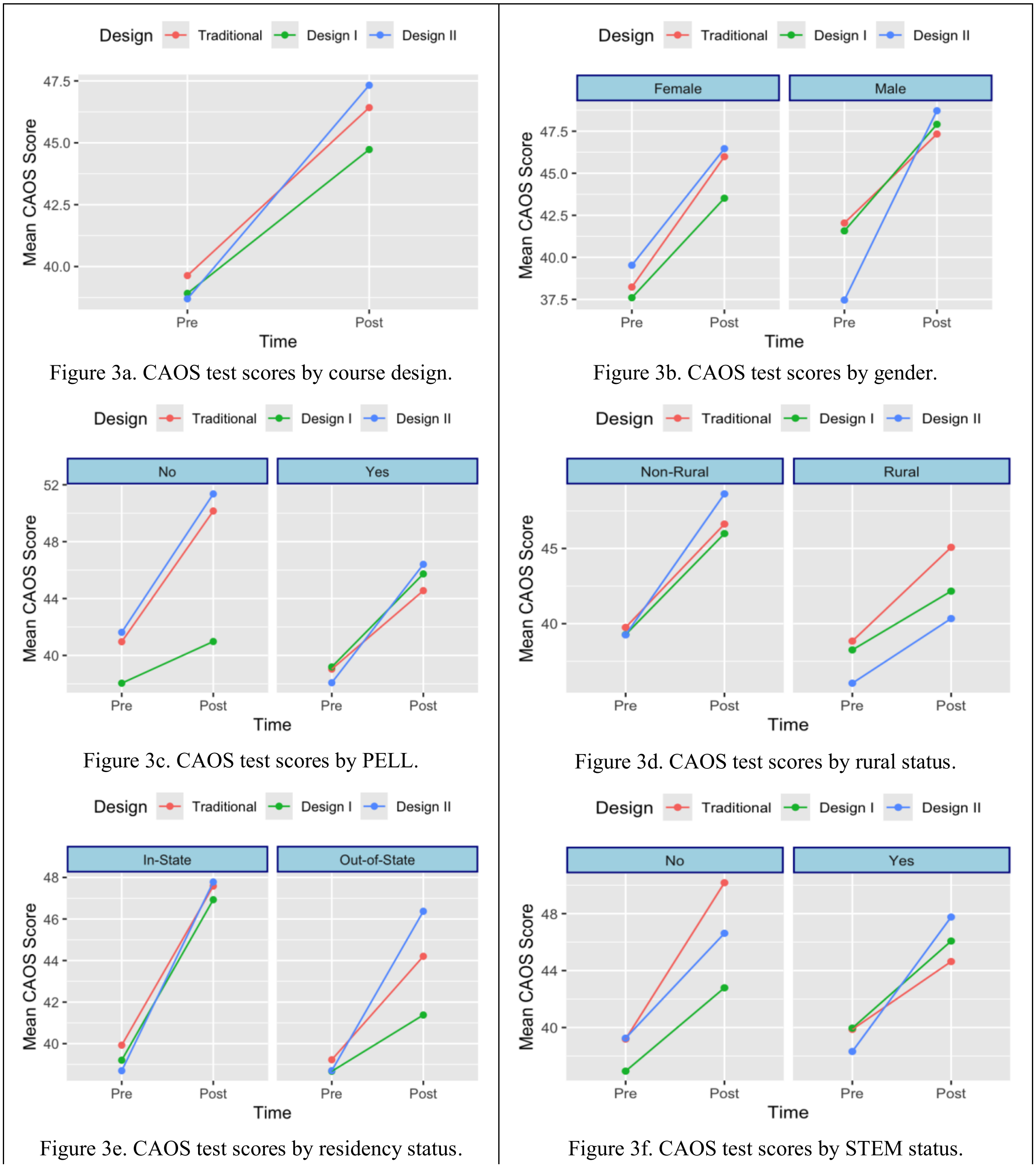

Figure 3a. CAOS test scores by course design.

Figure 3b. CAOS test scores by gender.

Figure 3c. CAOS test scores by PELL.

Figure 3d. CAOS test scores by rural status.

Figure 3e. CAOS test scores by residency status.

Figure 3f. CAOS test scores by STEM status.

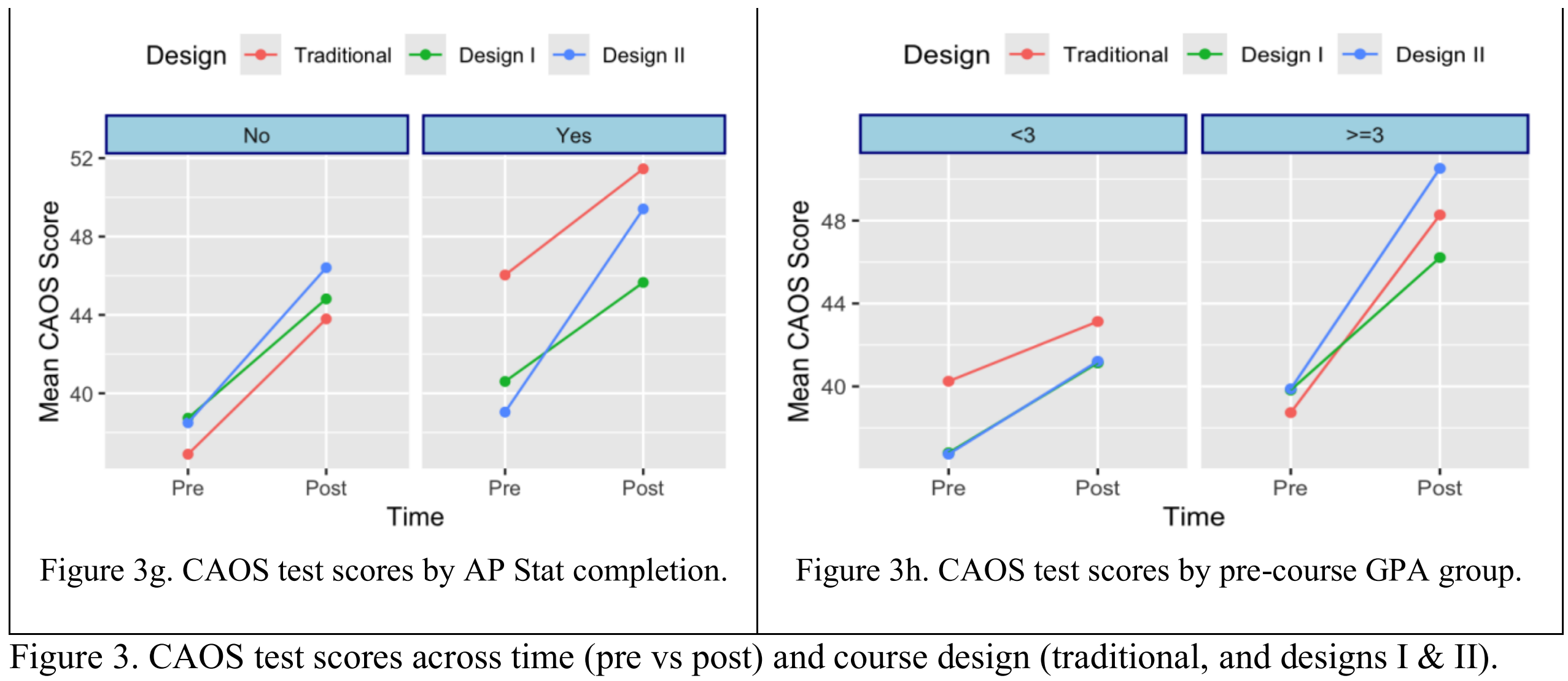

Figure 3g. CAOS test scores by AP Stat completion.

Figure 3h. CAOS test scores by pre-course GPA group.

Figure 3. CAOS test scores across time (pre vs post) and course design (traditional, and designs I & II).

- Gender: Both male and female students improved across all designs. Notably, females in Design II saw their scores rise from below 39.0 to 46.5, while males in the same group achieved the highest overall post-test average (48.0).
- Socioeconomic Status: Non-PELL students showed the largest conceptual gains, particularly in Design II (52.0). However, PELL recipients in Design II and Design I also showed strong improvement, reaching post-test scores of 46.0 to 47.0.
- STEM Status: Among STEM majors, Design II was the most effective format (48.0). Interestingly, non-STEM students in the Traditional sections achieved high post-test scores (50.0), though Design II also supported significant growth for this group (46.5).
- Prior Experience: Students without AP Statistics coursework benefited significantly from Design II, achieving a higher post-test average (46.0) than their counterparts in Design I (45.0) and the Traditional group (44.0).
- Overall: Design II sections exhibited the most consistent growth across nearly all subgroups.

*Statistical Shifts in Conceptual Mastery*

The results of partially overlapping t-tests indicate that all three instructional designs yielded statistically significant conceptual improvements ($p < .001$; Table 3, Panel A). Students in the Design II group exhibited the largest mean increase (9.31 points; $t = 4.65$, $p < .0001$), while those in Design I and Traditional sections showed gains of 6.73 and 6.82 points, respectively.

A two-way Type III repeated-measures ANOVA further revealed a statistically significant main effect of time, $F(1, 220) = 39.11$, $p < .0001$, GES = 0.064, confirming that CAOS scores increased significantly from pre-test to post-test across all instructional roles (Table 3, Panel B). In contrast, the interaction between Design × Time was not statistically significant ($F(2, 220) = 0.596$, $p = 0.552$, GES = 0.002), suggesting that the pattern of improvement over time was statistically comparable across the three formats in the omnibus model. The validity of these results was confirmed through an evaluation of model assumptions. Before conducting the ANOVA, the normality of the response variable was evaluated using Shapiro-Wilk tests and visual inspection of Q–Q plots. While Shapiro–Wilk tests indicated some departures from normality within specific groups, Q–Q plots suggested approximately symmetric distributions without severe deviations (see Appendix C). Given the moderate-to-large sample sizes in each condition and the established robustness of repeated-measures ANOVA to mild non-normality, the parametric model was retained. Furthermore, a non-parametric rank-based longitudinal analysis was conducted as a robustness check, yielding consistent results (see Appendix D, Table D1).

Table 3. Changes in CAOS performance by instructional design and time.

**Panel A**: Within-Design Pre–Post Change (Partially Overlapping t-tests)

| **Design** | **Mean Change** | **T-Statistic** | **P-value** |
|---|---|---|---|
| Traditional | 6.82 | 3.32 | 0.0008 |
| Design I | 6.73 | 3.66 | 0.0003 |
| Design II | 9.31 | 4.65 | <0.0001 |

**Panel B**: Repeated-Measures ANOVA (Between-Design Differences)

| **Effect** | **d.f.** | **GES** | **F-Statistic** | **P-value** |
|---|---|---|---|---|
| Design | 2, 220 | 0.005 | 0.849 | 0.4290 |
| Time | 1, 220 | 0.064 | 39.115 | <0.0001 |
| Design × Time | 2, 220 | 0.002 | 0.596 | 0.5520 |

**Note.** Panel A reports results from partially overlapping t-tests assessing within-design pre–post changes in CAOS scores (Derrick et al., 2017). Panel B reports Type III repeated-measures ANOVA results. GES = generalized eta squared, a measure of effect size.

*Predictors of Conceptual Mastery*

To identify the specific drivers of post-test performance on the CAOS assessment of conceptual understanding, we conducted a multivariable linear regression using MICE to address missing data (50 imputations, seed = 2026). The results are summarized in Table 4.

Table 4. Results of multivariable linear regression for the CAOS post-test score with MICE imputation.

| **Term** | | **Estimate** | **SE** | **95% LCL**[a] | **95% UCL**[a] | **P-value** |
|---|---|---|---|---|---|---|
| Intercept | | 38.46 | 6.763 | 25.07 | 51.84 | <0.0001 |
| Pretest Score | | 0.19 | 0.082 | 0.02 | 0.35 | **0.0242** |
| Design[b] | Design I | -0.62 | 2.250 | -5.06 | 3.82 | 0.7827 |
| | Design II | 0.68 | 1.901 | -3.07 | 4.42 | 0.7217 |
| Gender | Male | 1.74 | 1.834 | -1.87 | 5.36 | 0.3427 |
| PELL | Yes | -0.26 | 2.222 | -4.69 | 4.12 | 0.9065 |
| Rural | Yes | -3.08 | 2.386 | -7.79 | 1.62 | 0.1981 |
| Residency | Out-of-State | -2.78 | 1.875 | -6.47 | 0.92 | 0.1396 |
| STEM | Yes | -0.02 | 1.802 | -3.58 | 3.53 | 0.9896 |
| AP STAT | Yes | 1.12 | 2.077 | -2.99 | 5.22 | 0.5915 |
| Pre-Course GPA | ≥ 3.0 | 3.58 | 2.010 | -0.39 | 7.54 | 0.0765 |
| Attendance Rate | | -0.06 | 0.067 | -0.19 | 0.07 | 0.3504 |
| Course Grade | A | 7.32 | 2.845 | 0.86 | 13.80 | **0.0268** |
| | B | 5.68 | 2.820 | 0.12 | 11.24 | **0.0454** |
| | C | 2.63 | 2.616 | -2.54 | 7.80 | 0.3160 |
| **$R^2$ = 15.53%** | | **Adjusted $R^2$ = 11.32%** | | | **P-value**[c] **= 0.0533** | |

[a]LCL and UCL are the lower confidence limit and the upper confidence limit, respectively.
[b]The reference category for *Design* is "Traditional", *Gender* is "Female", *PELL* is "No", *Rural* is "No", *STEM* is "No", *AP STAT* is "No", *Pre-Course GPA* is "<3.0", and *Course grade* is "D".
[c]p-value was obtained from an F-test with degrees of freedom 14 and 281.

After adjusting for other variables in the model, initial knowledge (pre-test score) and final course grade emerged as the only significant predictors of conceptual mastery (post-test scores). Specifically, pre-test scores were positively associated with post-test scores ($\hat{\beta}$ = 0.19, $p$ = 0.024) and students earning an "A" scored significantly higher than those earning a "D" ($\hat{\beta}$ = 7.32, $p$ = 0.027), as did students earning a "B" ($\hat{\beta}$ = 5.68, $p$ = 0.045). In contrast, with imputation, demographic factors (Gender, PELL, Rurality, Residency), STEM status, prior academic preparation (AP Stat, GPA), and Course Design were not significant predictors of the CAOS post-test score. Nonetheless, the overall model was marginally statistically significant based on the pooled F-test (p = 0.053), and the model explained only 15.53% of the variance in post-test scores with an adjusted $R^2$ of 11.32%.

To ensure the robustness of these findings to missing data handling, a sensitivity analysis was conducted by comparing the MICE results against a complete-case analysis (see Appendix E, Table E1). While the direction and magnitude of most coefficients remained consistent, several predictors, specifically rural status and residency, that appeared significant in the complete-case analysis were no longer significant after imputation. Given that complete-case analysis assumes data are missing completely at random (MCAR), whereas multiple imputation relies on the less restrictive missing at random (MAR) assumption, the MICE results are considered more reliable. This suggests that the apparent effects of geography in simpler models were likely artifacts of non-random missing data rather than true drivers of statistical learning.

## 4.2. Readiness for Data Science

The second research question (RQ2) examines the impact of the VSCL on students' self-assessment of their data science competencies. As detailed in Table 1, DS readiness was measured using a series of survey items on a 6-point Likert scale ranging from strongly disagree (0) to strongly agree (5). Responses were averaged to create a single composite DS readiness score (min = 0, max = 5). The reliability and construct validity of the DS readiness measure were evaluated using both classical and model-based approaches in pre- and post-survey samples. Internal consistency was assessed using Cronbach's alpha ($\alpha$) and ordered McDonald's omega ($\omega$; Flora, 2020), while convergent validity was evaluated using average variance extracted (AVE) derived from confirmatory factor analysis (CFA). The pre-survey scale demonstrated strong internal consistency ($\alpha$ = 0.91; $\omega$ = 0.90) and adequate convergent validity (AVE = 0.71). Post-survey results indicated further metric improvement ($\alpha$ = 0.92; $\omega$ = 0.92; AVE = 0.74), confirming that the instrument provided a stable and valid measure of the readiness construct across the study period.

### *Trends and Subgroup Variations in DS Readiness*

Descriptive results indicate that while all students reported increased DS readiness over the course, the magnitude of growth was heavily dependent on instructional design (Figure 4a). Design II consistently led to the greatest gains, with post-test scores reaching approximately 4.0 (out of 5.0), followed by Design I (nearly 3.4). In contrast, the Traditional group showed the least progress, with post-test scores remaining below 2.7. Furthermore, analysis across demographic and academic subgroups (Figures 4b–4h) revealed that the VSCL formats were effective equalizers:

- Gender and Socioeconomic Status: Both male and female students in Design II reached readiness levels near 4.0, while the Traditional group scores remained stagnant between 2.5 and 2.8. Similarly, both PELL and non-PELL students benefited most from Design II.
- STEM Status: Notably, non-STEM students saw the most improvement in Design II, eventually reaching the highest readiness scores of any subgroup. This suggests the interactive labs were particularly effective at demystifying data analysis for students outside of quantitative majors.
- Prior Preparation: Students without prior AP Statistics coursework experienced the largest growth in Design II, whereas the Traditional format had a minimal negative impact on their perceived readiness.

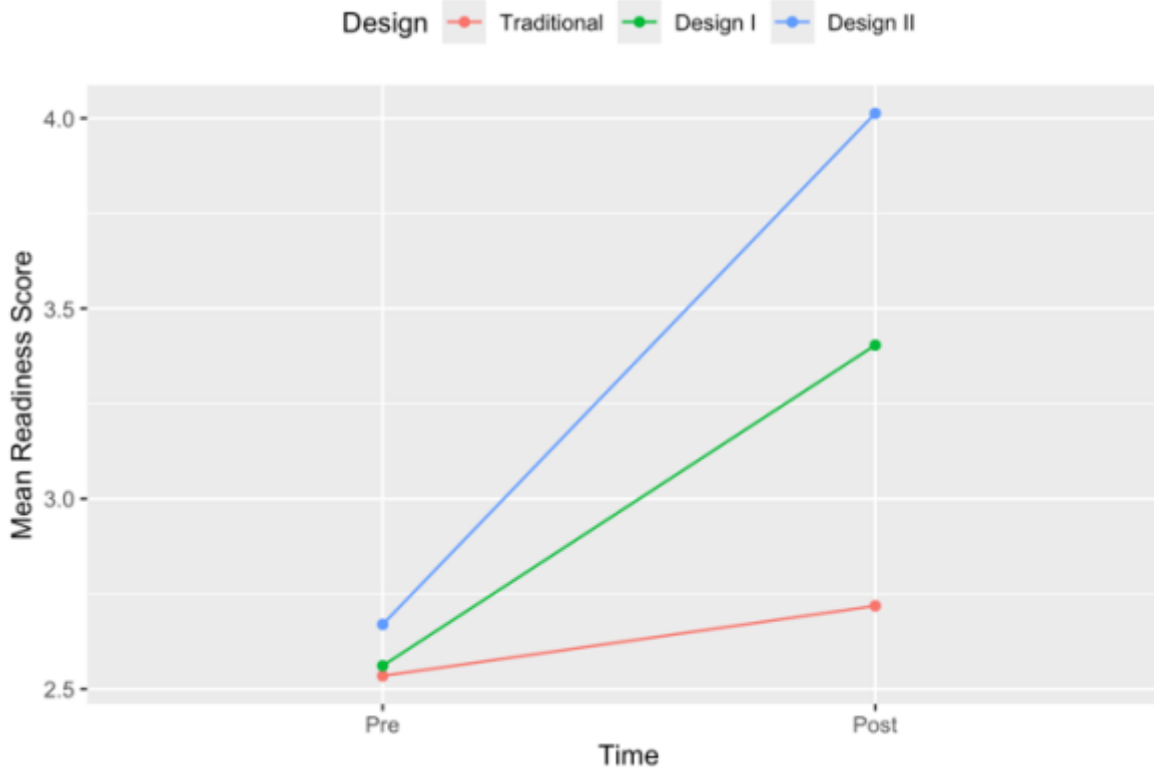

Figure 4a. DS readiness scores by course design.

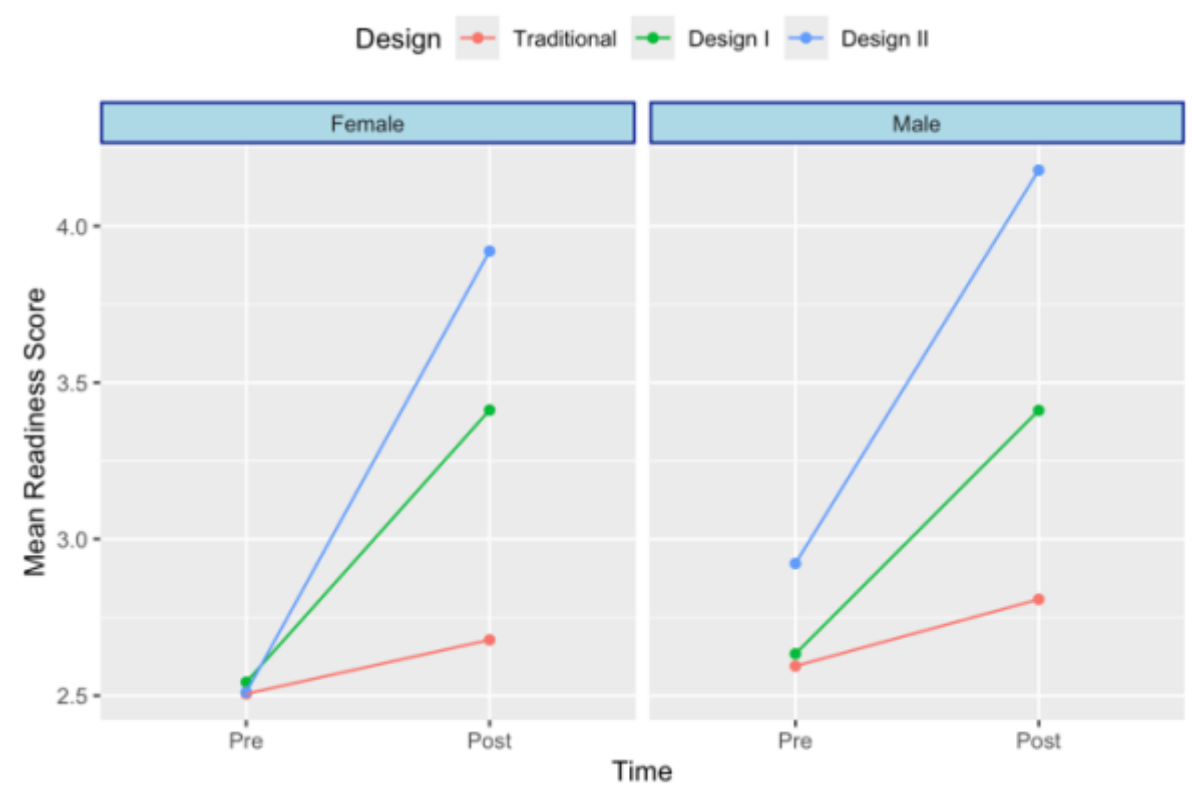

Figure 4b. DS readiness scores by gender.

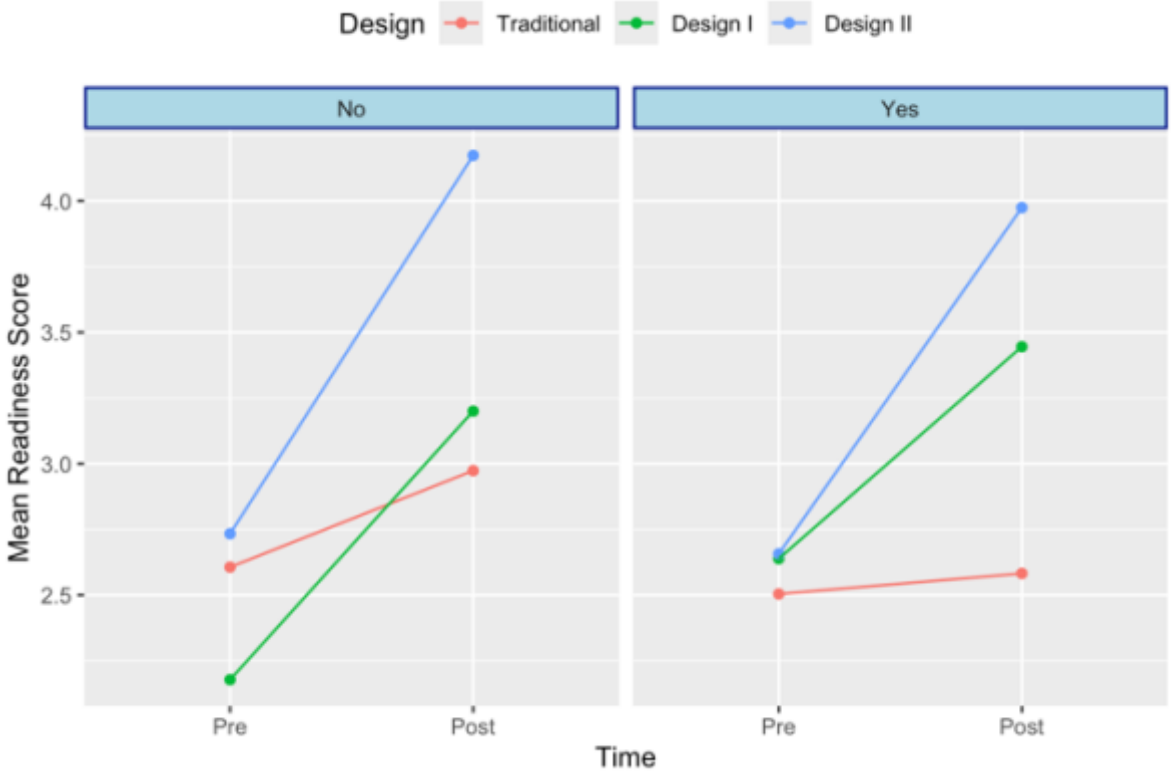

Figure 4c. DS readiness scores by PELL status.

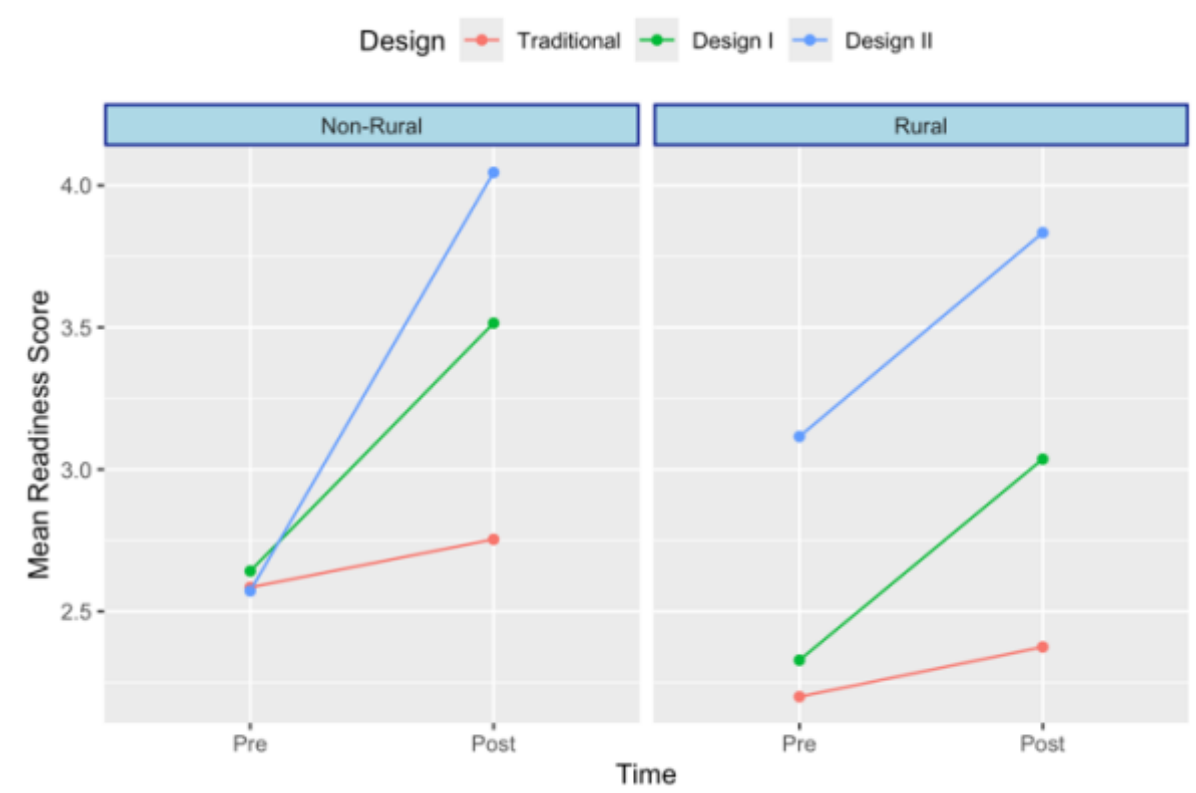

Figure 4d. DS readiness scores by rural status.

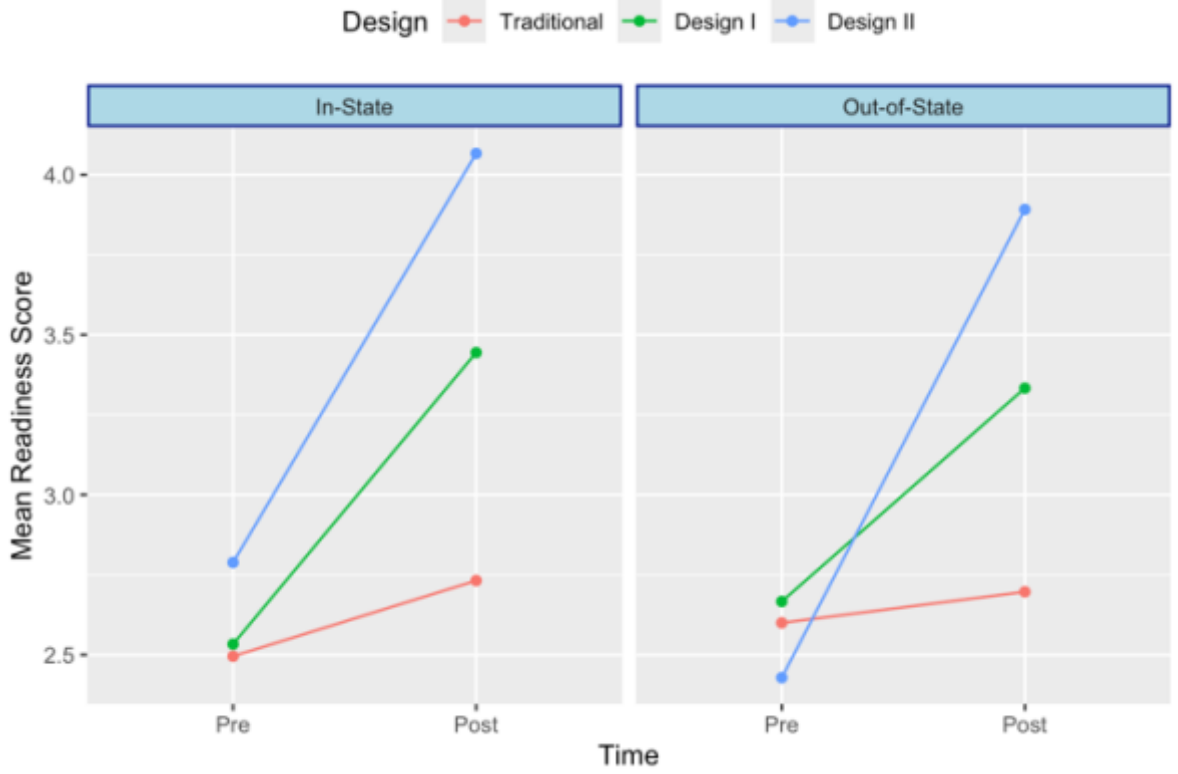

Figure 4e. DS readiness scores by residency.

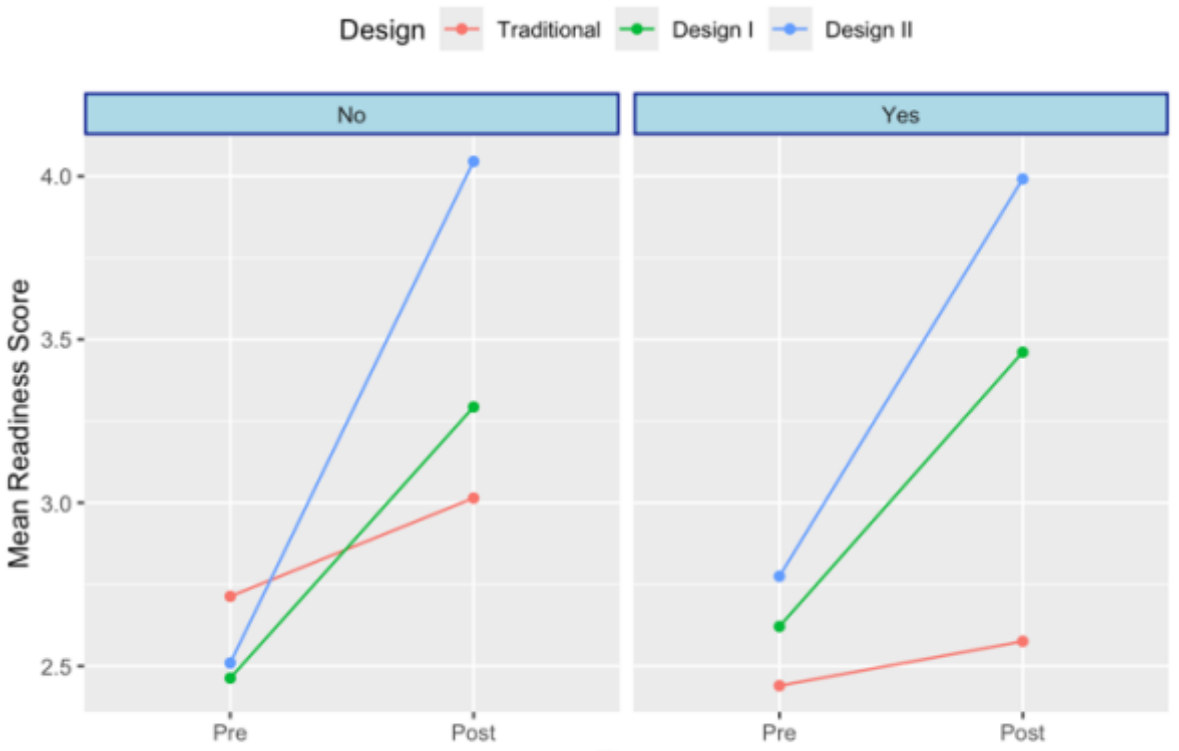

Figure 4f. DS readiness scores by STEM status.

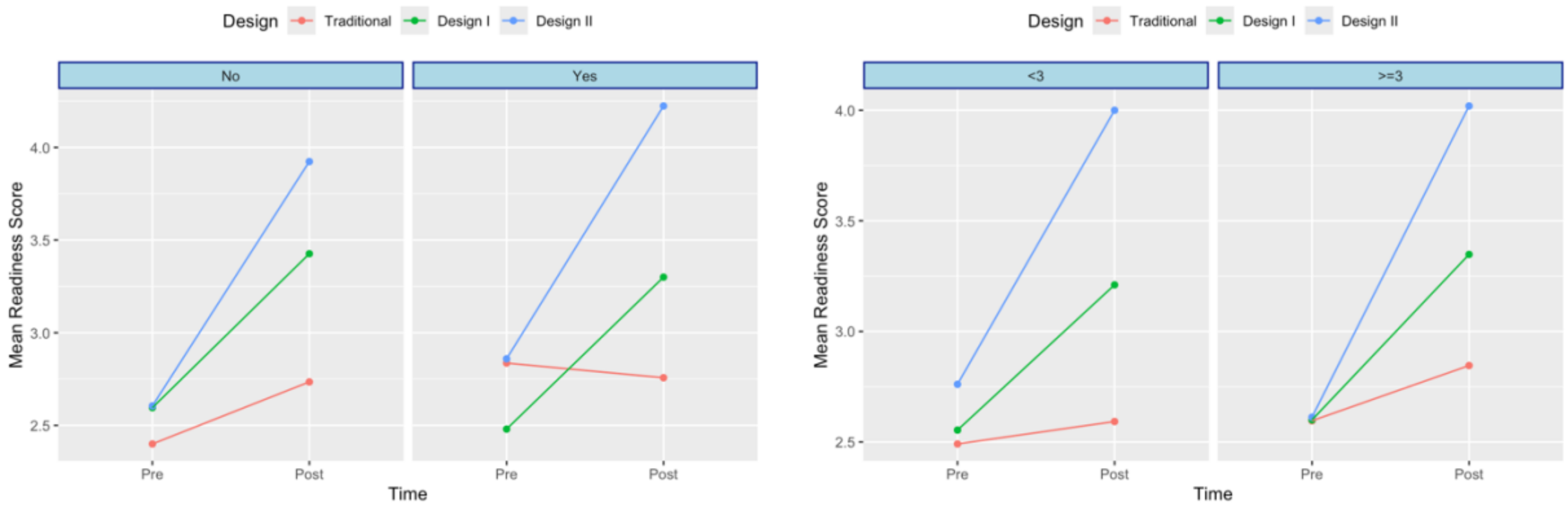

Figure 4g. DS readiness scores by AP stat completion.

Figure 4h. DS readiness scores by GPA group.

Figure 4. DS readiness across time (pre vs post) and course design (traditional, and designs I & II).

*Statistical Shifts in DS Readiness*

The results of partially overlapping t-tests confirm that the DS readiness gains among the VSCL groups were both substantial and highly significant (Table 5, Panel A). While the Traditional group showed only a marginal, non-significant gain (mean change = 0.18, $p = 0.090$), Design I (0.84, $p < .0001$) and Design II (1.34, $p < .0001$) showed significant statistical growth. The two-way Type III repeated-measures ANOVA revealed a highly significant Design × Time interaction, $F(2, 197) = 12.73$, $p < .0001$, GES = 0.049 (Table 5, Panel B). This confirms that the pattern of improvement in DS readiness was not uniform across groups; rather, the VSCL interventions significantly outperformed the traditional format in building student confidence in data analysis. The ANOVA model assumptions diagnostics showed minor departures from normality, but the moderate-to-large sample size justified the use of parametric ANOVA testing, with the non-parametric checks in Appendix D (Table D2) providing consistent results.

Table 5. Changes in DS readiness by instructional design and time.

**Panel A**: Within-Design Pre–Post Change (Partially Overlapping t-tests)

| **Design** | Mean Change | T-Test | P-value |
|---|---|---|---|
| Traditional | 0.18 | 1.35 | 0.0901 |
| Design I | 0.84 | 4.68 | <0.0001 |
| Design II | 1.34 | 9.54 | <0.0001 |

**Panel B**: Repeated-Measures ANOVA (Between-Design Differences)

| **Effect** | d.f. | GES | F-Test | P-value |
|---|---|---|---|---|
| Design | 2, 197 | 0.078 | 13.922 | <0.0001 |
| Time | 1, 197 | 0.114 | 63.520 | <0.0001 |
| Design × Time | 2, 197 | 0.049 | 12.732 | <0.0001 |

**Note.** Panel A reports results from partially overlapping t-tests assessing within-design pre–post changes in CAOS scores (Derrick et al., 2017). Panel B reports Type III repeated-measures ANOVA results. GES = generalized eta squared, a measure of effect size.

*Predictors of DS Readiness*

Multivariable linear regression using MICE imputation (50 imputations, seed = 2026) was employed to identify the primary drivers of post-course readiness (Table 6). The overall model was highly significant ($p$

< .0001) and explained a substantial portion of the variance in post-test DS readiness scores ($R^2$ = 31.53%; $R^2_{adj}$ = 28.12%). After controlling for all baseline characteristics, instructional design was the strongest predictor of DS readiness. Students in Design I scored significantly higher than the Traditional group ($\hat{\beta}$ = 0.72, $p$ = 0.0004), and students in Design II reported even greater gains ($\hat{\beta}$ = 1.28, $p$ < .0001). Pre-test scores were also positively associated with final readiness scores ($\hat{\beta}$ = 0.19, $p$ = 0.0056). Notably, demographic factors, including gender, PELL status, and STEM major, were not significant predictors, suggesting that the curriculum design, rather than student background, was the primary driver of confidence.

A sensitivity analysis comparing the MICE results with results from complete-case analysis confirmed the stability of these findings (see Appendix E, Table E2). The MICE-based model provided narrower coefficient confidence intervals, justifying the focus on the imputed results under the MAR assumption.

Table 6. Results of multivariable linear regression for the DS readiness post-test scores with MICE imputation.

| **Term** | | **Estimate** | **SE** | **95% LCL[a]** | **95% UCL[a]** | **P-value** |
|---|---|---|---|---|---|---|
| Intercept | | 1.88 | 0.501 | 0.89 | 2.87 | 0.0003 |
| Pretest Score | | 0.19 | 0.069 | 0.06 | 0.33 | **0.0056** |
| Design[b] | Design I | 0.72 | 0.196 | 0.35 | 1.11 | **0.0004** |
| | Design II | 1.28 | 0.160 | 0.96 | 1.60 | **<0.0001** |
| Gender | Male | 0.07 | 0.150 | -0.23 | 0.36 | 0.6467 |
| PELL | Yes | -0.05 | 0.186 | -0.42 | 0.32 | 0.7856 |
| Rural | Yes | -0.35 | 0.206 | -0.76 | 0.06 | 0.0932 |
| Residency | Out-of-State | -0.06 | 0.159 | -0.38 | 0.25 | 0.7026 |
| STEM | Yes | -0.10 | 0.158 | -0.41 | 0.21 | 0.5367 |
| AP STAT | Yes | -0.06 | 0.165 | -0.39 | 0.27 | 0.7131 |
| Pre-Course GPA | ≥3.0 | -0.01 | 0.171 | -0.35 | 0.33 | 0.5972 |
| Attendance Rate | | 0.004 | 0.005 | -0.01 | 0.01 | 0.6133 |
| Course Grade | A | 0.30 | 0.276 | -0.25 | 0.84 | 0.2862 |
| | B | 0.08 | 0.266 | -0.45 | 0.61 | 0.7613 |
| | C | 0.19 | 0.223 | -0.26 | 0.63 | 0.4075 |
| $R^2$ **= 31.53%** | | **Adjusted** $R^2$ **= 28.12%** | | | **P-value[c] < 0.0001** | |

[a]LCL and UCL are the lower and upper confidence limits, respectively.
[b]The reference category for *Design* is "Traditional", *Gender* is "Female", *PELL* is "No", *Rural* is "No", *STEM* is "No", *AP STAT* is "No", *Pre-Course GPA* is "<3.0", and *Course grade* is "D".
[c]p-value was obtained from an F-test with 14 and 281 degrees of freedom.

## 4.3. Data Science Aspirations

The third research question (RQ3) evaluates the impact of integrating computing (R coding) into introductory statistics on students' aspirations of data science. Unlike the previous scales, DS aspirations were analyzed separately rather than as a composite score. Each item captures a distinct level of students' academic intentions related to data science, ranging from individual coursework to degree completion.

Table 7 compares DS aspirations across the three instructional formats, specifically among the subsample of students who reported prior awareness of the field. Overall, the proportion of students planning to take additional DS courses showed a slight downward trend across all cohorts. In the Traditional group, interest declined from 16.07% to 13.79% ($N$ = 58). More pronounced decreases were observed in the experimental groups: Design I dropped from 23.53% to 9.30%, and Design II declined from 16.07% to 8.62%. Aspirations for formal credentials (certificates, minors, or degrees) remained low throughout the semester. Certificate pursuit remained under 9% across all formats, with Design I showing the highest post-course interest (6.98%). In terms of DS minor aspirations, Design I saw the most notable growth in this category, rising from 0.00% at the pre-survey to 6.98% at the post-survey. Interest in a full undergraduate DS degree remained limited, with slight increases in the Traditional (3.45%) and Design I (4.65%) groups, while Design II saw interest dropping from one student to none.

Table 7. DS aspirations among students who have ever heard about DS grouped by course design.

| | | N of Yes / Total N* (% of Yes) | | |
|---|---|---|---|---|
| **Question** | **Design** | **Traditional** | **Design I** | **Design II** |
| Plan to take DS courses? | Presurvey | 9 / 56 (16.07) | 8 / 34 (23.53) | 9 / 56 (16.07) |
| | Postsurvey | 8 / 58 (13.79) | 4 / 43 (9.30) | 5 / 58 (8.62) |
| Plan to complete a UG DS Certificate? | Presurvey | 2 (3.57) | 3 (8.82) | 2 (3.57) |
| | Postsurvey | 3 (5.17) | 3 (6.98) | 1 (1.72) |
| Plan to complete a UG DS minor? | Presurvey | 1 (1.79) | 0 (0.00) | 0 (0.00) |
| | Postsurvey | 1 (1.72) | 3 (6.98) | 0 (0.00) |
| Plan to complete a UG DS Degree? | Presurvey | 1 (1.79) | 0 (0.00) | 1 (1.79) |
| | Postsurvey | 2 (3.45) | 2 (4.65) | 0 (0.00) |

*Total *N* for the certificate, minor, and degree questions is the same as for the courses question.

Table 8. DS course aspirations across designs grouped by gender, STEM status, and AP Stat completion.

| | | N of Yes / Total N (%) of Yes | | | | | |
|---|---|---|---|---|---|---|---|
| **Question** | **Design** | **Traditional** | | **Design I** | | **Design II** | |
| | | Male | Female | Male | Female | Male | Female |
| Do you plan to take a DS course? | Presurvey | 6/22 (27.3) | 3/34 (8.8) | 3/15 (20.0) | 5/19 (26.3) | 3/26 (11.5) | 6/30 (20.0) |
| | Postsurvey | 3/19 (15.8) | 5/39 (12.8) | 3/15 (20.0) | 1/27 (3.7) | 1/20 (5.0) | 4/38 (10.5) |
| | | STEM | Non-STEM | STEM | Non-STEM | STEM | Non-STEM |
| | Presurvey | 6/37 (16.2) | 3/19 (15.8) | 8/26 (30.8) | 0/8 (0.0) | 5/34 (14.7) | 4/22 (18.2) |
| | Postsurvey | 3/38 (7.9) | 5/20 (25.0) | 3/31 (9.7) | 1/11 (9.1) | 2/13 (15.4) | 2/25 (8.0) |
| | | AP STAT | No-AP STAT | AP STAT | No-AP STAT | AP STAT | No-AP STAT |
| | Presurvey | 3/18 (16.7) | 6/38 (15.8) | 3/12 (25.0) | 5/22 (22.7) | 3/15 (20.0) | 6/41 (14.6) |
| | Postsurvey | 3/22 (13.6) | 5/34 (14.7) | 2/10 (20.0) | 2/32 (6.3) | 2/13 (15.4) | 3/25 (7.3) |

Table 8 provides a more granular view of aspirations to take future DS coursework, disaggregated by gender, STEM status, and prior AP Statistics experience, across course formats. The gender-based results reveal diverging patterns across the three instructional formats. In the Traditional group, male course aspirations declined (27.3% to 15.8%) while female aspirations showed a slight increase (8.8% to 12.8%). In the experimental groups, however, female aspirations saw sharper declines: in Design I, female aspirations dropped from 26.3% to 3.7%, and in Design II, they decreased from 20.0% to 10.5%. Initially, STEM students in Design I expressed the highest level of interest (30.8%), though this dropped significantly to 9.7% post-course. Conversely, the Traditional group saw a post-course shift among non-STEM students, whose interest increased from 15.8% to 25.0%. When considering prior preparation, students with AP Statistics experience consistently reported higher baseline intentions across all designs. While interest generally waned by the end of the term, students with AP credit in Design II proved more resilient, showing a smaller decline (from 20.0% to 15.4%) compared to their peers without prior AP exposure (from 14.6% to 7.3%).

# 5. DISCUSSION

## 5.1. Conclusions

This study examined the impact of integrating R coding via a Virtual Statistical Computing Lab (VSCL) into an introductory statistics course, contrasting three instructional formats—Traditional (no computing lab), Design I (Posit Cloud-based static lab), and Design II (*learnr*-based interactive lab). We evaluated outcomes across three areas: conceptual learning gains, data science readiness, and data science aspirations.

### *Conceptual Learning Gains and Pedagogical Drivers*

All instructional formats yielded statistically significant improvement on the CAOS test from pre- to post-course. Design II students exhibited the greatest gains (mean improvement = 9.31 points), followed by Design I (6.73 points) and the Traditional group (6.07 points). These results suggest that integrating R computing labs, particularly interactive *learnr*-based labs, can facilitate enhanced students' conceptual understanding of statistics. The superior performance of Design II points to specific pedagogical elements that differentiate it from the static environment. While Design I provided a uniform environment via Posit Cloud, addressing the logistical "double wall" identified by Doehler and Taylor (2015), Design II added layers of immediate feedback and scaffolded checks for understanding. According to student feedback collected by the project's external evaluator (see Appendix F), these embedded R code activities and guided tutorials were perceived as the most beneficial for conceptual growth. By providing an interactive sandbox where code and concepts are unified, Design II likely reduced the cognitive load associated with syntax, allowing students to focus on the algorithmic intuition, as defined by Kaplan (2007), required for statistical reasoning. Multivariable regression indicated that, controlling for pretest score and other demographic and academic factors, earning a grade of B or higher in the course had significant positive associations with post-test scores. Course design, though not statistically significant in the full model, showed a trend favoring the interactive format. These findings align with prior literature suggesting that computing-enriched curricula support conceptual growth, especially among academically prepared students (Tintle et al., 2015; Horton & Hardin, 2021).

### *Data Science Readiness and Confidence*

Students in both Design I and Design II reported significantly greater gains in self-perceived readiness for data science (DS) tasks compared to the Traditional group. This is statistically substantiated by the significant Time × Design interaction in our ANOVA model, which confirms that the trajectory of student confidence was fundamentally different across the three formats. While the Traditional group's readiness scores remained stagnant, the experimental cohorts showed a diverging, upward trend. This suggests that the active ingredients of the VSCL, such as immediate feedback and reproducible workflows, were the primary drivers of growth. Regression analysis further confirmed instructional design as a significant independent predictor of post-course DS readiness, even after adjusting for demographic and academic covariates. These results support research emphasizing the importance of real-time feedback and embedded computing experiences in building student confidence with DS tools and workflows (Beckman et al., 2021; Hardin et al., 2021; Stoudt et al., 2022). Notably, Design II consistently outperformed Design I in readiness across various subgroups, including PELL-eligible students, those from rural backgrounds, and non-STEM majors. This echoes studies highlighting the advantages of interactive computing in promoting deep engagement (Woodard & Lee, 2021). The lab's ability to provide a low-stakes entry point into coding appears to be the primary mechanism for narrowing the confidence gap for students typically underrepresented in data-intensive fields.

*The Aspiration Gap: Skill vs. Belonging*

Student aspirations to pursue DS coursework or credentials were generally low and declined slightly by the end of the course. In Design II, the percentage of students planning to take DS courses fell from 16.07% to 8.62%, and interest in completing a DS certificate dropped from 3.57% to 1.72%. These results contrast with prior findings suggesting that early exposure to statistical computing enhances STEM motivation (Freeman et al., 2014; Gnanadesikan et al., 2014). This decline in aspirations may reflect external factors such as a lack of visibility into DS career paths/perceived inaccessibility of the field among minority students, or a "reality check" phenomenon where students gain a more accurate, and perhaps daunting, understanding of the rigor involved in data wrangling and reproducible scripting. However, student feedback collected by the project's external evaluator (see Appendix F) provides nuance into which elements influenced students who remained interested in a data science path. While the Fall 2022 cohort identified specific datasets and informational posts on careers as most influential, the Spring 2023 cohort cited technical engagement, specifically the virtual computing labs and R Markdown project reports, as the primary drivers. This suggests that while career-contextualized data is vital for sparking interest, the hands-on experience of navigating a professional computing environment provides the "active ingredients" necessary to entice students to persist despite the field's perceived rigor.

## 5.2. Implications

The findings underscore the value of integrating interactive R-based computing into introductory statistics, not only for boosting conceptual understanding but also for significantly increasing students' self-confidence in data analysis. To transition students from a state of "readiness" (I can do this) to one of "aspiration" (I want to be this), these results suggest several instructional adjustments. First, lab-based career contextualization should be enhanced by moving beyond technical checkpoints to include "professional identity" modules that use authentic datasets to mimic the workflows of data analysts in fields relevant to minority-serving institutions, such as public health or regional economics. Second, to address the "data science person" perception and bridge the gap between skill development and long-term intent, virtual labs should incorporate complementary interventions like intentional mentoring and Career Spotlight videos featuring diverse practitioners to help students visualize their own future in the field. Finally, because Design II fostered the highest levels of readiness, it should be utilized as a strategic recruitment tool; by creating clear curricular bridges and explicitly demonstrating how *learnr* skills translate into second-level coursework, instructors may mitigate the observed decline in aspirations by reducing the perceived difficulty of advancing in the discipline.

## 5.3. Limitations

Several limitations should be considered when interpreting the study's findings. First, while the overall sample provided sufficient power for primary analyses, the sample sizes for individual instructional formats were relatively small, particularly for Design I. This reduction in group-level N restricted our ability to detect more nuanced interactions within demographic and academic subgroups. Furthermore, while we controlled for GPA and prior AP Statistics experience, unmeasured variables such as instructor-specific pedagogical styles, section-specific peer dynamics, and time-of-day effects remain potential confounding factors. Second, the unique academic background and timing of the study must be acknowledged. The participants consisted largely of early-career college students whose secondary education was disrupted by the COVID-19 pandemic. This disruption likely influenced their initial technical preparedness and may have heightened the "fear of syntax" described by Doehler and Taylor (2015). Furthermore, as is characteristic of many minority-serving institutions, this course serves a high proportion of non-STEM majors fulfilling a terminal mathematics requirement. For these students, high levels of pre-existing math anxiety often frame statistics as a graduation hurdle rather than a professional pathway. This suggests that the baseline for data science aspirations in this population may be lower and less malleable within a single semester than in courses populated by STEM-focused majors at research-intensive universities. Finally, measurement constraints may affect the generalizability of the results. Our assessment of data science readiness and aspirations relied on self-report surveys, which are inherently subject to social desirability bias. Additionally, because the CAOS test and post-course surveys were administered for participation credit rather than as high-stakes assessments, the effort expended by some students may not reflect their maximum potential performance. Despite these constraints, this study provides valuable early insights into the utility of virtual R-based labs and offers a rigorous foundation for scaling these interventions in diverse educational settings.

## 5.4. Future Directions

This study offers preliminary but important evidence on the effectiveness of integrating interactive R-based computing labs into introductory statistics instruction. Further research should examine whether participation in this instructional model influences students' future academic trajectories, particularly in data science. Tracking enrollment in follow-on courses, changes in declared majors, and persistence in STEM pathways could shed light on longer-term outcomes. Additionally, future research should explore how to translate gains in confidence and readiness into concrete aspirations. Supplementing technical instruction with exposure to real-world applications, career panels, or alumni testimonials could help students envision themselves in data-intensive fields. Finally, piloting the VSCL model across diverse institutions can help refine content and pedagogy, especially when supported by randomized section designs and larger, more diverse student populations. Ultimately, building data science capacity among undergraduates requires more than access to computing tools—it demands integrated, equitable, and engaging instructional approaches that promote both competence and belonging.

## Acknowledgement

This work was supported by National Science Foundation grant # EHR 2106945.

## Appendix A: Illustrations of lab and class material for the DS-Infused Intro Stats course

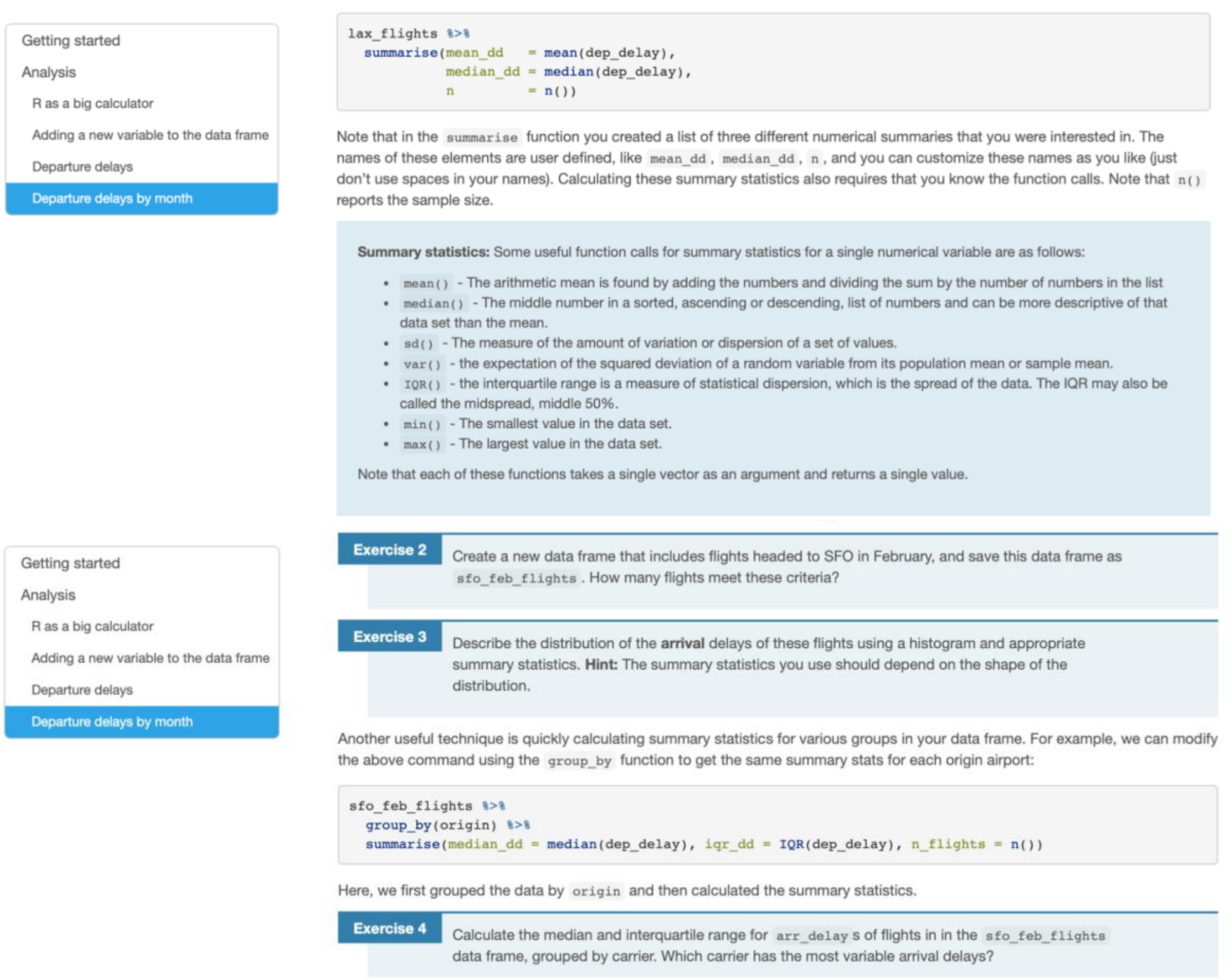

Figure A1: Example static computing lab demo and exercises under design I (VSCL via Posit Cloud).

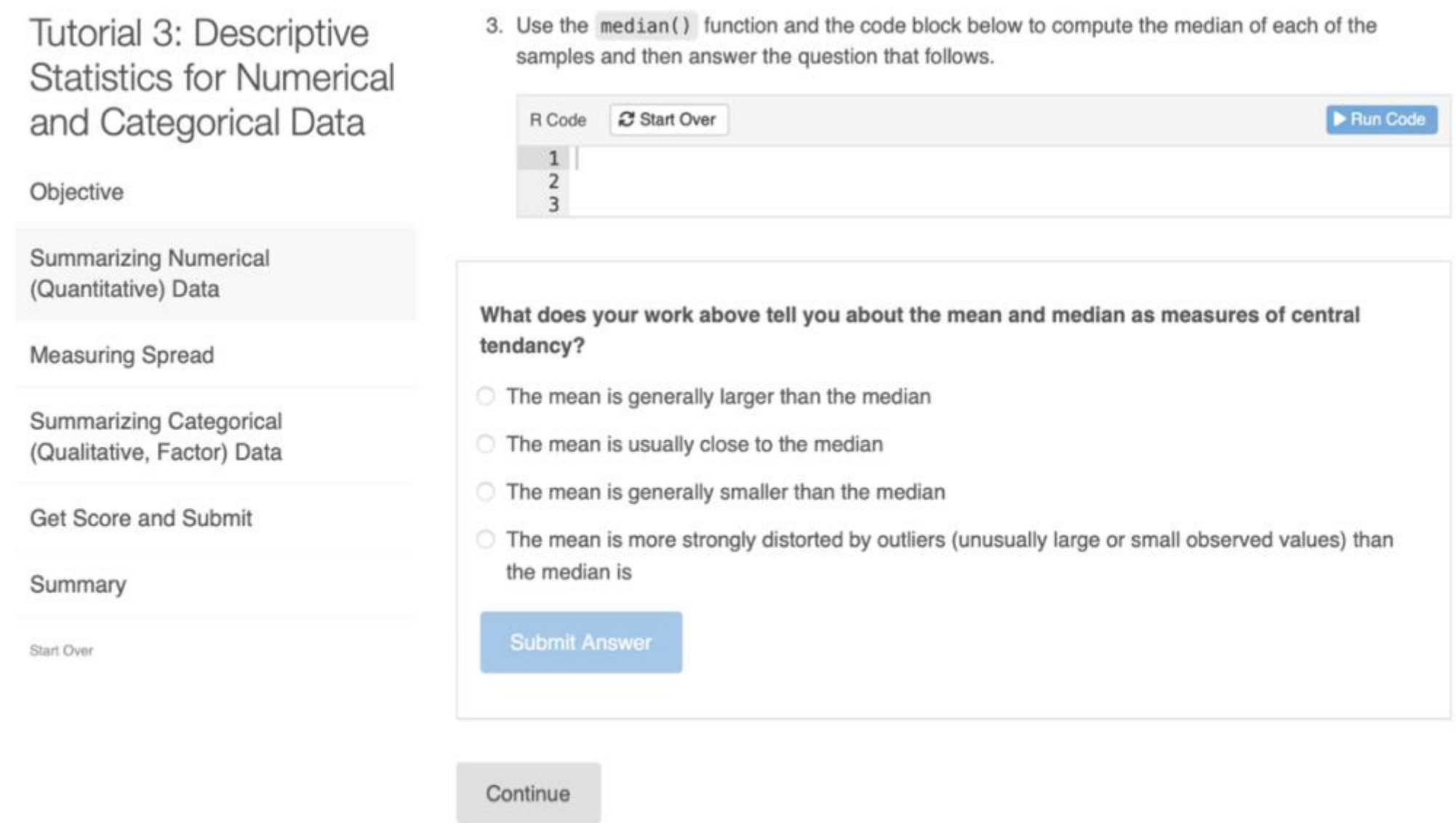

Figure A2: Excerpt from an R Shiny pre-lab tutorial used for both course designs (I & II).

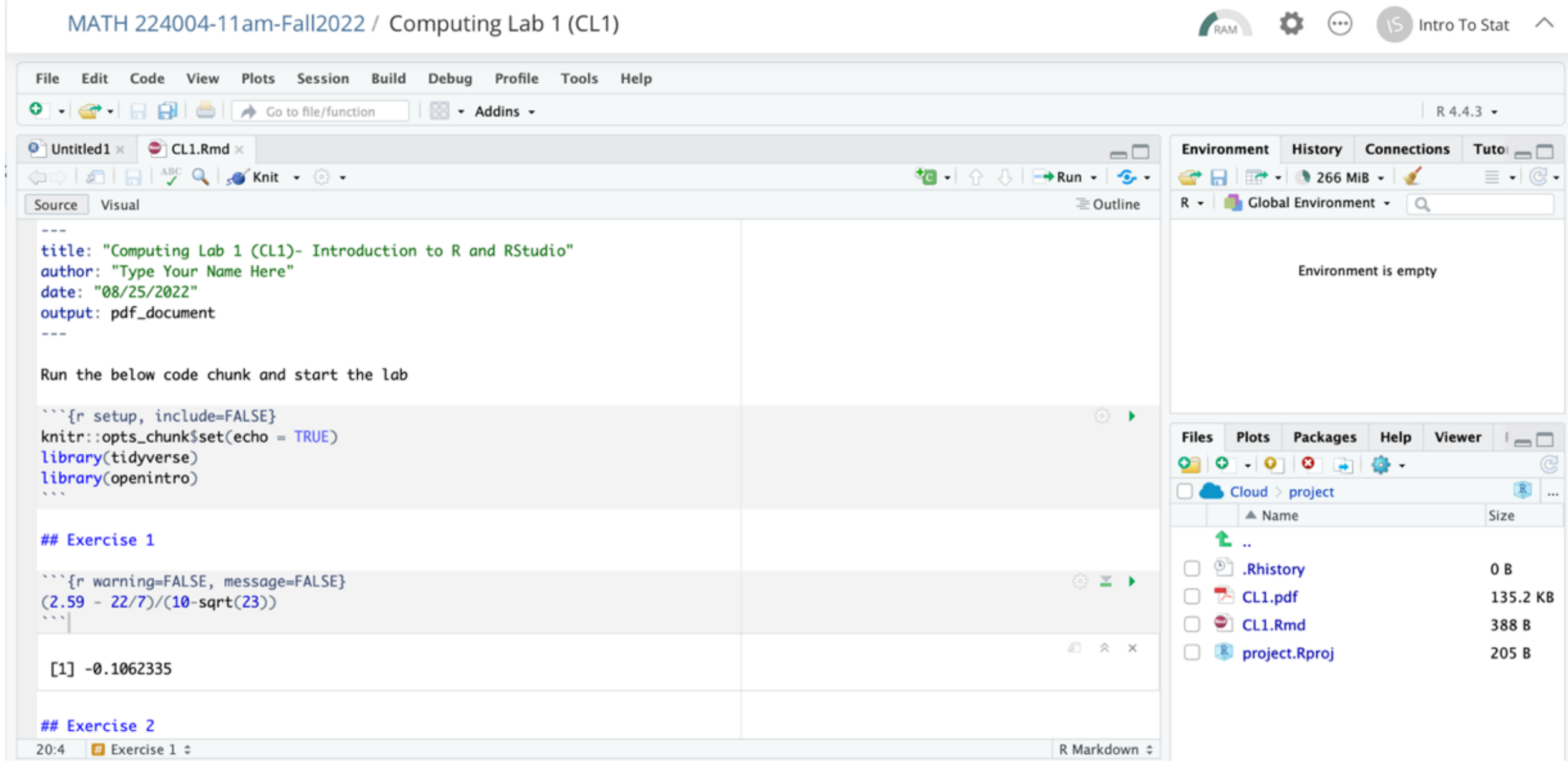

Figure A3: Computing lab R markdown report template under design I (VSCL via Posit Cloud).

## Exploratory Data Analysis Part I

Start Over

Recall that the five number summary includes the min, first quantile (Q1), median, third quantile (Q3), and max. Using the `mpg` dataset, we can compute the five number summary of the vehicle's highway mileage `hwy` as follows.

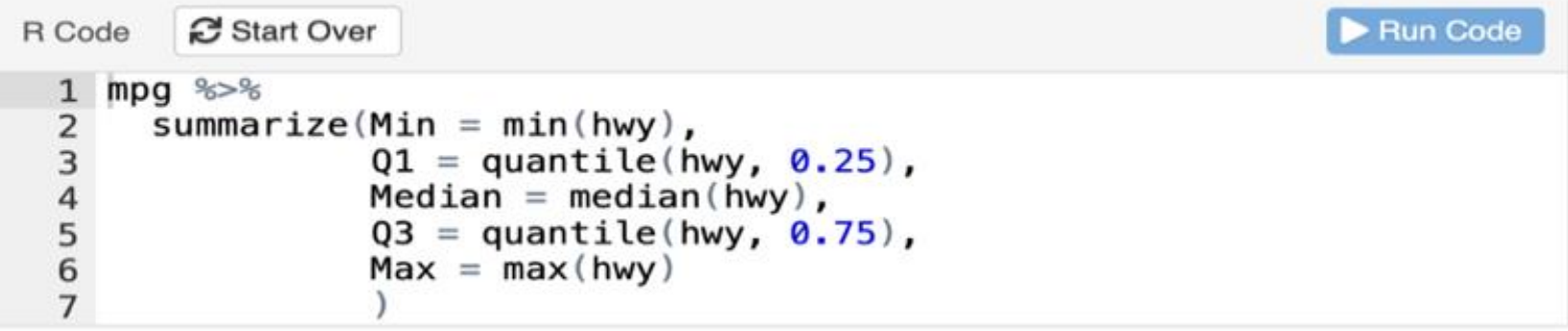

Notice how the `quantile()` function is used to obtain quantiles by setting the proportion of data below the quantile (i.e., 0.25 or 0.75)

4. Use the code chunk below to calculate the measures of center (mean and median) for the vehicle's city mileage `cty`.

R Code  Start Over  Run Code  Submit Answer

1
2
3

5. Use the code chunk below to calculate the variation measures (standard deviation and interquartile range) for the vehicle's city mileage `cty`.

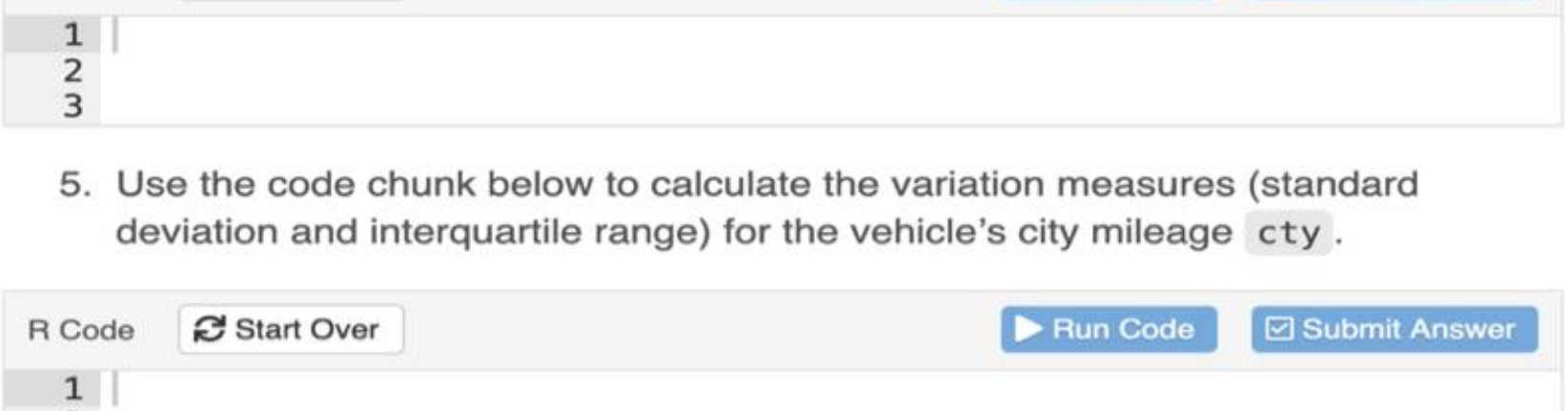

Figure A4: Example interactive computing lab demo and exercises under design II (VSCL via *learnr*).

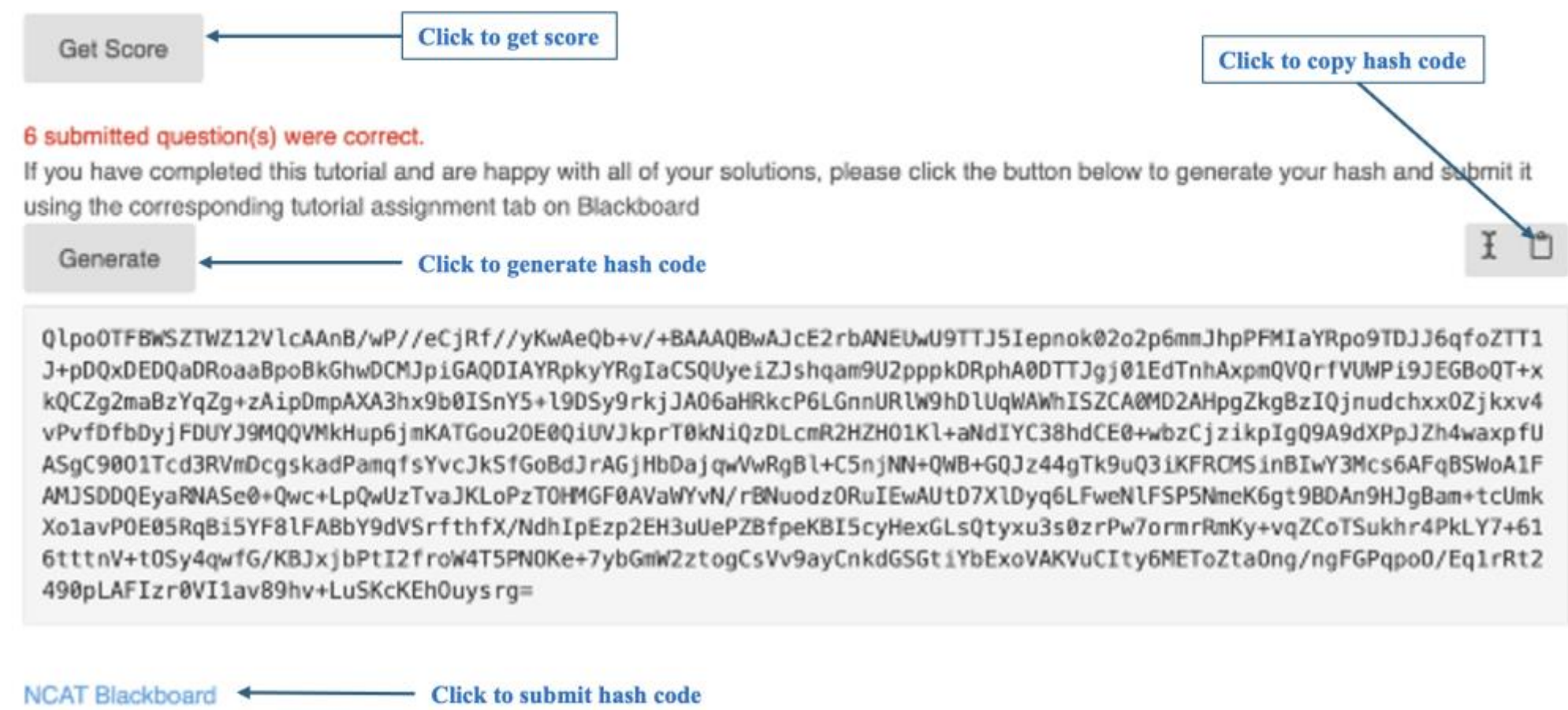

Figure A5: Computing lab submission process under design II (VSCL via *learnr*).

```r
library(learnr)
library(gradethis)
library(learnrhash)
library(tidyverse)

listfile <- list.files(pattern = "txt",full.names = T, recursive = TRUE)
n = length(listfile)
dat = c()
score = c()

for(i in 1:n){
  x1 = file(listfile[i],"r")
  r_line = readLines(x1)
  name = sub("\\).*", "", sub(".*\\(", "", r_line[1])) #Reads the first line and picks up the name
  hash = sub("\\.*</pre>", "", sub(".*\\id=\"hash_output\">", "", r_line[7])) #Reads the hash submission
  temp_dat = c(name, hash)
  dat = rbind(dat, temp_dat)
}

row.names(dat) = NULL
colnames(dat) = c("Username","hash")
dat = as.data.frame(dat)

for(i in 1:nrow(dat)) {
  score[i] = learnrhash::decode_obj(dat$hash[i]) %>%
    filter(correct == "TRUE") %>%
    summarise(score = round(n()/10*20, 2)) %>%
    pull(score)
}

dat = cbind(dat, score)
dat = dat %>% select(Username, score)

CL1 <- read_csv("gc_MATH224001.202320_studinfo_2023-01-21-19-26-31.csv")
CL1 = left_join(CL1, dat, by = "Username")
CL1 = CL1%>%
  mutate(`CL1 [Total Pts: 10 Score]` = score)%>%
  select(-c("Student ID","Last Access","Availability","score"))

write.csv(CL1, "MATH224001_CL1.csv", row.names = F)
```

Figure A6: R script for processing students' lab submissions under design II (VSCL via *learnr*).

## Examples

**Example 1.** Calculate the mean of a sample with five observations: 5, 3, 8, 5, 6.

$$\bar{x} = \frac{\sum_{i=1}^{n} x_i}{n} = \frac{5+3+8+5+6}{5} = \frac{27}{5} = 5.4$$

Using R, we can calculate the mean using the `mean()` command. Notice that we need to put the values in a vector using the `c()` function which stands for *concatenate*.

R Code | Start Over | Run Code

```
1 mean(c(5,3,8,5,6))
2
3
```

Figure A7: Example lecture slide with interactive coding used for design II (VSCL via *learnr*).

## Appendix B: Partially Overlapping Test Statistic Details

The test-statistic for partially overlapping tests, as defined by Derrick et al. (2017), is

$$t = \frac{(\bar{x}_1 - \bar{x}_2)}{\sqrt{\frac{s_1^2}{n_1} + \frac{s_2^2}{n_2} - 2r\frac{s_1 s_2 n_c}{n_1 n_2}}},$$

where $\bar{x}_1$ and $\bar{x}_2$ ($s_1^2$ and $s_2^2$) are the sample means (variances) from all available observations in sample 1 (pre-test scores) and sample 2 (post-test scores), respectively, $r$ is the Pearson correlation coefficient for paired observations only (i.e., observations with both pre- and post-test scores available), $n_c$ is the number of paired observations, and $n_1$ and $n_2$ are the total number of observations in sample 1 and sample 2, respectively.

## Appendix C: Two-Way Repeated Measures ANOVA Assumptions Diagnostics

*CAOS Test Scores*

Table C1: Summary statistics for the CAOS test score by course design and time.

| Role | Time | N | Mean | SD |
|---|---|---|---|---|
| Traditional | Pre | 67 | 40.75 | 11.145 |
| Traditional | Post | 67 | 47.97 | 14.217 |
| Design I | Pre | 57 | 39.45 | 11.274 |
| Design I | Post | 57 | 44.61 | 12.349 |
| Design II | Pre | 99 | 39.18 | 11.523 |
| Design II | Post | 99 | 47.20 | 14.613 |

Table C2: Outlier detection for the CAOS test score by course design and time.

| Role | Time | Respondent ID | Score | Outlier | Extreme |
|---|---|---|---|---|---|
| Traditional | Pre | 220277 | 78.79 | Yes | No |
| Design I | Post | 220226 | 77.88 | Yes | No |
| Design I | Post | 220313 | 78.79 | Yes | No |
| Design II | Pre | 230107 | 72.73 | Yes | No |
| Design II | Pre | 230179 | 72.73 | Yes | No |
| Design II | Pre | 230224 | 72.73 | Yes | No |
| Design II | Pre | 230225 | 75.76 | Yes | No |

Table C3: Normality test using the Shapiro-Wilk Test for the CAOS test score by course design.

| Role | Time | Statistic (W) | P-value |
|---|---|---|---|
| Traditional | Pre | 0.960 | 0.0317 |
| Traditional | Post | 0.977 | 0.2610 |
| Design I | Pre | 0.980 | 0.4460 |
| Design I | Post | 0.937 | 0.0053 |
| Design II | Pre | 0.922 | <0.001 |
| Design II | Post | 0.955 | 0.0021 |

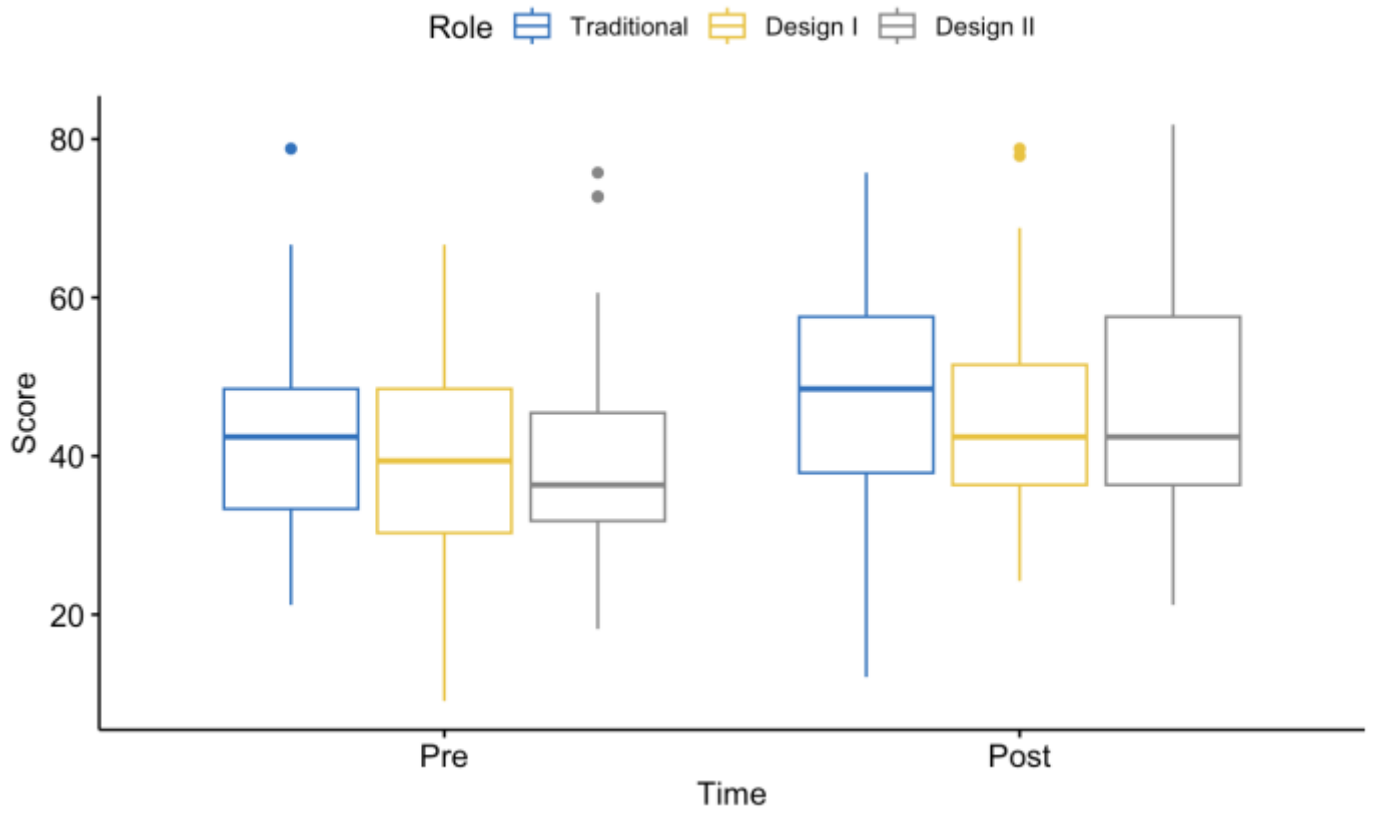

**Figure C1.** Boxplots of CAOS pre- and post-test scores by course design

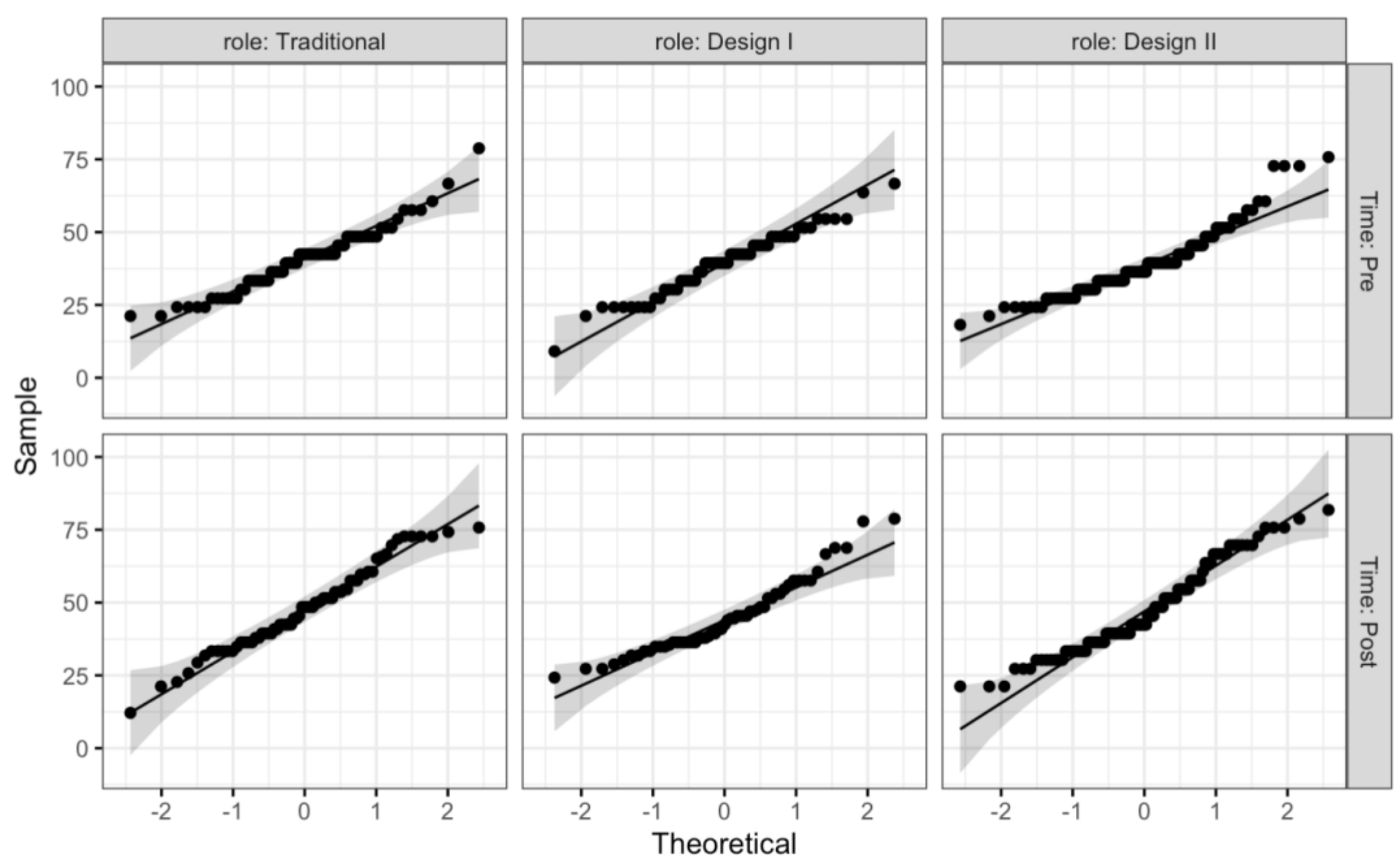

**Figure C2.** Q–Q plots assessing normality of CAOS scores by course design and time.

*DS Readiness Scores*

Table C4: Summary statistics for the DS readiness score by design and time.

| Role | Time | N | Mean | SD |
|---|---|---|---|---|
| Traditional | Pre | 80 | 2.52 | 1.07 |
| Traditional | Post | 80 | 2.74 | 1.07 |
| Design I | Pre | 48 | 2.63 | 1.05 |
| Design I | Post | 48 | 3.40 | 1.12 |
| Design II | Pre | 72 | 2.67 | 1.11 |
| Design II | Post | 72 | 3.99 | 0.96 |

Table C5: Outlier detection for the DS readiness post-test score.

| Role | Time | Respondent ID | Score | Outlier | Extreme |
|---|---|---|---|---|---|
| Design II | Post | 230170 | 1 | Yes | No |
| Design II | Post | 230202 | 1 | Yes | No |

Table C6: Normality test using the Shapiro-Wilk Test for the DS readiness score by course design.

| Role | Time | Statistic (W) | P-value |
|---|---|---|---|
| Traditional | Pre | 0.956 | 0.0073 |
| Traditional | Post | 0.955 | 0.0070 |
| Design I | Pre | 0.941 | 0.0179 |
| Design I | Post | 0.965 | 0.1570 |
| Design II | Pre | 0.938 | 0.0016 |
| Design II | Post | 0.907 | <0.001 |

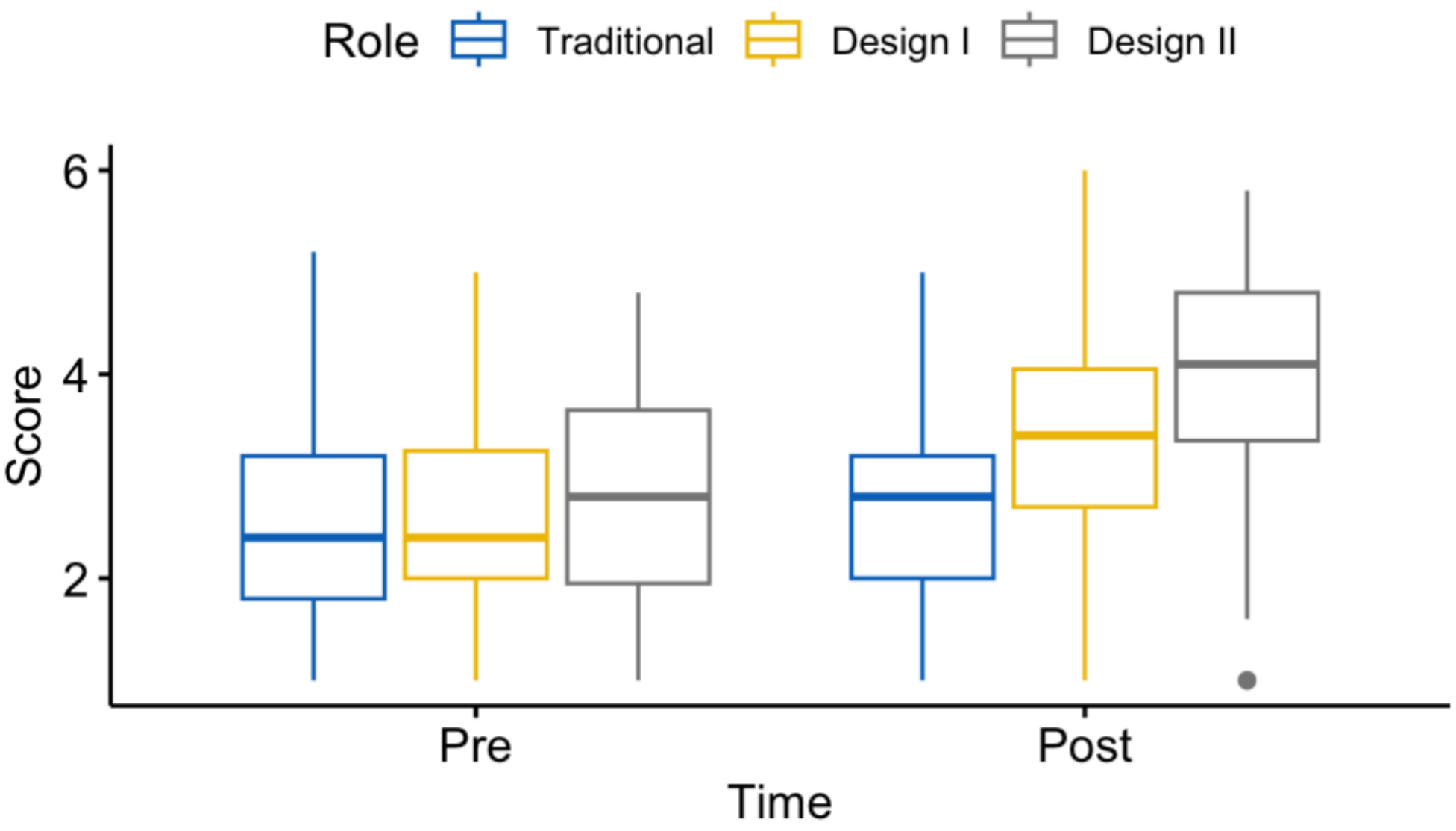

**Figure C3.** Boxplots of DS Readiness pre- and post-test scores by course design.

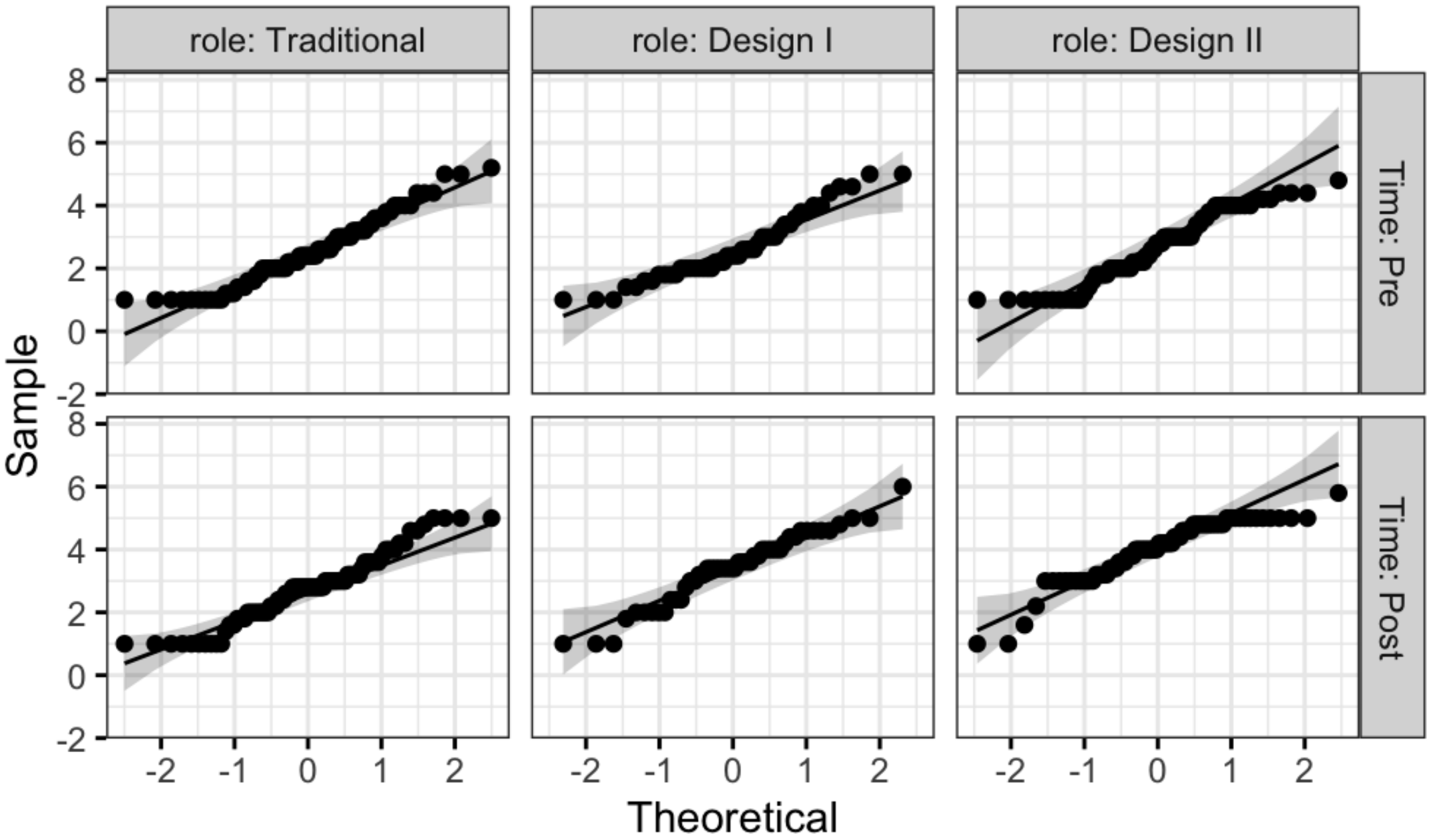

**Figure C4.** Q–Q plots assessing normality of DS readiness scores by course design and time.

## Appendix D: Nonparametric Repeated Measures ANOVA Results

Table D1: Non-parametric ANOVA results for the CAOS test score.

| Effect | Statistic | df | p-value |
|---|---|---|---|
| Group | 1.08 | 1.938 | 0.3383 |
| Time | 32.73 | 1.000 | <0.0001 |
| Group × Time | 0.95 | 1.991 | 0.3854 |

Table D2: Non-parametric ANOVA results for the DS readiness score.

| Effect | Statistic | df | p-value |
|---|---|---|---|
| Group | 11.83 | 1.899 | <0.0001 |
| Time | 66.12 | 1.000 | <0.0001 |
| Group × Time | 10.87 | 1.962 | <0.0001 |

## Appendix E: Complete-Case-Analysis Regression Results

Table E1. Results of multivariable linear regression for the CAOS post-test score from complete case analysis.

| Term | | Estimate | SE | 95% LCL[a] | 95% UCL[a] | P-value |
|---|---|---|---|---|---|---|
| Intercept | | 33.73 | 8.639 | 16.686 | 50.780 | 0.0001 |
| Pretest Score | | 0.19 | 0.088 | 0.021 | 0.368 | **0.0283** |
| Design[b] | Design I | -0.98 | 2.801 | -6.506 | 4.547 | 0.7270 |
| | Design II | -1.09 | 2.421 | -5.866 | 3.689 | 0.6536 |
| Gender | Male | 1.90 | 2.160 | -2.359 | 6.165 | 0.3795 |
| PELL | Yes | 1.09 | 2.780 | -4.395 | 6.578 | 0.6951 |
| Rural | Yes | -6.01 | 2.999 | -11.931 | -0.095 | **0.0465** |
| Residency | Out-of-State | -6.18 | 2.411 | -10.936 | -1.419 | **0.0112** |
| STEM | Yes | -0.25 | 2.133 | -4.463 | 3.955 | 0.9052 |
| AP STAT | Yes | 1.94 | 2.330 | -2.658 | 6.539 | 0.4060 |
| Pre-Course GPA | ≥ 3.0 | 4.84 | 2.393 | 0.114 | 9.558 | **0.0448** |
| Attendance Rate | | -0.01 | 0.084 | -0.174 | 0.159 | 0.9297 |
| Course Grade | A | 6.68 | 3.924 | -1.061 | 14.425 | 0.0903 |
| | B | 7.77 | 3.304 | 1.250 | 14.290 | **0.0198** |
| | C | 3.21 | 3.145 | -2.999 | 9.412 | 0.3093 |
| **$R^2$ = 21.34%** | | **Adjusted $R^2$ = 15.19%** | | | **P-value[c] < 0.0001** | |

[a]LCL and UCL are the lower confidence limit and the upper confidence limit, respectively.
[b]Reference category for *Design* is “Traditional”, *Gender* is “Female”, *PELL* is “No”, *Rural* is “No”, *STEM* is “No”, *AP STAT* is “No”, *Pre-Course GPA* is “<3.0”, and *Course grade* is “D”.
[c]p-value was obtained from an F-test with degrees of freedom 14 and 179.

Table E2. Results of multivariable linear regression for the DS readiness post-test score from complete case analysis.

| Term | | Estimate | SE | 95% LCL[a] | 95% UCL[a] | P-value |
|---|---|---|---|---|---|---|
| Intercept | | 1.63 | 0.578 | 0.49 | 2.77 | 0.0055 |
| Pretest Score | | 0.22 | 0.072 | 0.08 | 0.36 | **0.0029** |
| Design[b] | Design I | 0.52 | 0.212 | 0.10 | 0.94 | **0.0147** |
| | Design II | 1.19 | 0.181 | 0.83 | 1.55 | **<0.0001** |
| Gender | Male | 0.12 | 0.176 | -0.23 | 0.47 | 0.4899 |
| PELL | Yes | -0.07 | 0.225 | -0.51 | 0.38 | 0.7618 |
| Rural | Yes | -0.34 | 0.250 | -0.84 | 0.15 | 0.1723 |
| Residency | Out-of-State | 0.02 | 0.188 | -0.35 | 0.39 | 0.9067 |
| STEM | Yes | -0.05 | 0.171 | -0.39 | 0.28 | 0.7539 |
| AP STAT | Yes | -0.12 | 0.188 | -0.49 | 0.25 | 0.5302 |
| Pre-Course GPA | ≥3.0 | -0.08 | 0.183 | -0.44 | 0.28 | 0.6503 |
| Attendance Rate | | 0.01 | 0.006 | -0.01 | 0.02 | 0.2560 |
| Course Grade | A | 0.30 | 0.306 | -0.30 | 0.91 | 0.3250 |
| | B | 0.10 | 0.267 | -0.42 | 0.63 | 0.7015 |

| C | 0.19 | 0.241 | -0.28 | 0.67 | 0.4245 |
|---|---|---|---|---|---|
| **$R^2$ = 28.38%** | | **Adjusted $R^2$ = 22.41%** | | **P-value[c] < 0.0001** | |

[a]LCL and UCL are the lower confidence limit and the upper confidence limit, respectively.
[b]The reference category for *Design* is "Traditional", *Gender* is "Female", *PELL* is "No", *Rural* is "No", *STEM* is "No", *AP STAT* is "No", *Pre-Course GPA* is "<3.0", and *Course grade* is "D".
[c]p-value was obtained from an F-test with degrees of freedom 14 and 168.

## Appendix F: Student Feedback Survey Summary

In each of Fall 2022 and Spring 2023, the project's external evaluator administered the Student Feedback Survey to students enrolled in treatment and control sections of the Intro Stats course to assess perceptions of course components. The survey was administered online during the last week of the semester.

### *Fall 2022*

In Fall 2022, a total of 108 students completed the survey (treatment = 51; control = 57). The survey was administered by the project's external evaluator following the conclusion of the semester.

#### Perceived Support for Understanding Statistical Concepts

Students in the treatment condition ranked eight course components from most (1) to least (8) beneficial for supporting their understanding of statistical concepts. Mean rankings and frequencies of most and least beneficial ratings are presented in Table F1. The OpenIntro textbook ($M$ = 3.06), interactive tutorials ($M$ = 3.18), and the virtual computing lab (R/RStudio; $M$ = 4.03) were ranked as the most beneficial instructional components.

Table F1. Ranked beneficial in understanding statistical concepts.

| Element | Average Rank | Most Beneficial | Least Beneficial |
|---|---|---|---|
| OpenIntro Textbook | 3.06 | 12 (33%) | 3 (8%) |
| Interactive tutorial (reading assignments) | 3.18 | 15 (40%) | 1 (3%) |
| Virtual computing lab (R/RStudio) | 4.03 | 5 (15%) | 3 (9%) |
| Information on data science jobs | 4.18 | 5 (15%) | 1 (3%) |
| Lab reports (R Markdown) | 4.23 | 2 (7%) | 0 |
| Datasets used | 4.60 | 2 (7%) | 3 (10%) |
| Information on data science educational opportunities | 4.76 | 2 (6%) | 5 (15%) |
| Data analysis project | 5.83 | 1 (3%) | 13 (43%) |

$n$ = 30-38; Scale: 1 = Most important to 8 = Least important

#### Perceived Influence on Data Science Aspirations

Students who indicated interest or uncertainty about pursuing data science ($n$ = 6–8) ranked the same course components according to their influence on career aspirations. Mean rankings and frequencies are presented in Table F2. Datasets used in the course ($M$ = 3.17) and information on data science careers ($M$ = 3.38) were perceived as the most influential components in shaping students' interest in data science.

Table F2. Ranked influential in enticing students to pursue data science.

| Element | Average Rank | Most Influential | Least Influential |
|---|---|---|---|
| Datasets used | 3.17 | 0 | 0 |
| Information on data science jobs | 3.38 | 2 (25%) | 0 |
| OpenIntro Textbook | 3.5 | 3 (50%) | 2 (33%) |
| Information on data science educational opportunities | 4.38 | 1 (13%) | 1 (13%) |
| Data analysis project | 4.57 | 1 (14%) | 2 (29%) |
| Virtual computing lab (R/RStudio) | 4.67 | 0 | 1 (17%) |
| Lab reports (R Markdown) | 5.17 | 0 | 0 |

| Interactive tutorial (reading assignments) | 5.33 | 1 (17%) | 0 |
|---|---|---|---|

*n* = 6-8; Scale: 1 = Most influential to 8 = Least influential

***Spring 2023***

In Spring 2023, a total of 107 students completed the Student Feedback Survey (treatment = 75; control = 32).

**Perceived Support for Understanding Statistical Concepts**

Students in the treatment condition ranked eight course components from most (1) to least (8) beneficial. Mean rankings and frequencies are presented in Table F3. Lecture slides with embedded interactive R code (*M* = 2.39), interactive tutorials (*M* = 2.96), and the virtual computing lab (R/RStudio; *M* = 3.44) were rated as the most beneficial components for supporting conceptual understanding.

Table F3. Ranked beneficial in understanding statistical concepts.

| Element | Average Rank | Most Beneficial | Least Beneficial |
|---|---|---|---|
| Lecture slides with interactive R codes | 2.39 | 29 (54%) | 4 (7%) |
| Interactive tutorial (reading assignments) | 2.96 | 8 (15%) | 2 (4%) |
| Virtual computing lab (R/RStudio) | 3.44 | 10 (19%) | 2 (4%) |
| Datasets used | 4.37 | 2 (4%) | 3 (6%) |
| OpenIntro Textbook | 4.48 | 5 (10%) | 6 (13%) |
| Data analysis project | 4.63 | 3 (6%) | 5 (9%) |
| Information on data science educational opportunities | 5.83 | 2 (4%) | 14 (30%) |
| Information on data science job | 6.38 | 1 (2%) | 9 (20%) |

*n* = 48-54; Scale: 1 = Most beneficial to 8 = Least beneficial

**Perceived Influence on Data Science Aspirations**

Students who expressed interest or uncertainty about pursuing data science (*n* = 6) ranked course components according to their influence on aspirations. Results are presented in Table F4. The virtual computing lab (*M* = 2.67) and R Markdown lab reports (*M* = 3.50) were perceived as the most influential elements in shaping interest in data science.

Table F4. Ranked influential in enticing students to pursue data science.

| Element | Average Rank | Most Influential | Least Influential |
|---|---|---|---|
| Virtual computing lab (R/RStudio) | 2.67 | 2 (33%) | 0 |
| Lab reports (R Markdown) | 3.50 | 1 (17%) | 0 |
| Data analysis project | 4.00 | 1 (17%) | 0 |
| Interactive tutorial (reading assignments) | 4.17 | 1 (17%) | 0 |
| Lecture slides with interactive R codes | 5.33 | 0 | 0 |
| Information on data science job | 5.50 | 1 (17%) | 0 |
| OpenIntro Textbook | 6.33 | 0 | 2 (33%) |
| Information on data science educational opportunities | 6.50 | 0 | 1 (17%) |
| Datasets used | 7.00 | 0 | 3 (50%) |

*n = 6*; Scale: 1 = Most influential to 9 = Least influential